\documentclass[twocolumn,showpacs,amsmath,amssymb,eqsecnum]{revtex4}

\usepackage{graphicx}

\newcommand{\cxconj}{\text{c.c.}}

\begin{document}

\title{%
Gauge-invariant Hamiltonian dynamics\\
of cylindrical gravitational waves%
}

\author{Ioannis Kouletsis}
\author{Petr H\'{a}j\'{\i}\v{c}ek}%
\affiliation{%
Institute of Theoretical Physics, University of Berne,\\
Berne, Switzerland%
}
\author{Ji\v{r}\'{\i} Bi\v{c}\'{a}k}
\affiliation{%
Institute of Theoretical Physics, Charles University,\\
V Hole\v{s}ovi\v{c}k\'{a}ch 2, 182 00 Prague 8,\\
Czech Republic\\
}

\date{August 11, 2003} 

\begin{abstract} The model of cylindrical gravitational waves is employed to
  work out and check a recent proposal in Ref.\ \cite{honnef} how a
  diffeomorphism-invariant Hamiltonian dynamics is to be constructed. The
  starting point is the action by Ashtekar and Pierri because it contains the
  boundary term that makes it differentiable for non-trivial variations at
  infinity. With the help of parametrization at infinity, the notion of gauge
  transformation is clearly separated from that of asymptotic symmetry.  The
  symplectic geometry of asymptotic symmetries and asymptotic time is
  described and the role of the asymptotic structures in defining a
  zero-motion frame for the Hamiltonian dynamics of Dirac observables is
  explained. Complete sets of Dirac observables associated with the asymptotic
  fields are found and the action of the asymptotic symmetries on them is
  calculated. The construction of the corresponding quantum theory is
  sketched: the Fock space, operators of asymptotic fields, the Hamiltonian
  and the scattering matrix are determined.
\end{abstract}

\pacs{04.20.Fy, 04.60.-m, 04.60.Ds, 04.60.Kz}

\maketitle

\section{Introduction}
A deep problem in quantum gravity is the dependence of the constructed quantum
theory on the choice of coordinates. A famous example of a quantization method
based on such a choice of gauge is the Arnowitt, Deser and Misner reduction
\cite{ADM}. There have been ideas of how this problem could be avoided, such
as Dirac's operator constraint equations \cite{dirac}, Bergmann's work on Dirac
observables \cite{bergmann}, BRST method \cite{HT}, Euclidean quantum gravity
\cite{HG} etc. None of these methods has as yet been really successful because
new problems always emerged.

In this paper, we focus on the method based on the Dirac observables. In a
sense, this is the most straightforward one: all variables that are used have
to be gauge invariant. However, changes of coordinates include changes of time
so that the gauge invariance entails the time independence: Dirac observables
must be integrals of motion. Two problems that are related to this have been
noticed already by Bergmann \cite{bergmann}: ``{\em frozen dynamics}'' and
{\em scarcity}: the dynamics of Dirac observables is trivial---they just stay
constant, and it is difficult to find {\em any} such quantity in general
relativity. More recently, the {\em non-locality} of such variables has been
proved \cite{torrobs}: the expression for any Dirac observable in terms of
local fields must contain derivatives of all orders.

As for the ``frozen dynamics'', Rovelli's idea of ``evolving constants of
motion'' \cite{rovelli} has shown one way out of the problem. Another, but not
completely unrelated way, is based on the observation that any Hamiltonian
formulation of dynamics needs a frame \cite{KT}. A possible frame for a
non-trivial Hamiltonian dynamics of Dirac observables has been specified in
\cite{JMP1} and \cite{DDO} under the assumption that there is a symmetry. In
\cite{honnef}, it is shown that even an asymptotic symmetry suffices, at least
for a simple finite-dimensional model.

Concerning the scarcity, there have been different proposals of how Dirac
observables could be constructed (e.g., \cite{argent}). Also, in the case
of asymptotically flat models, there are even complete sets of natural
constants of motion, namely the in- and out-fields. These quantities have been
described by DeWitt \cite{DW2} within his covariant perturbation theory and by
Ashtekar \cite{AA} in general.

Finally, the non-locality of the kind found need not always lead to a real
problem. For example, in any non-trivial quantum field theory in Minkowski
spacetime, the expressions for the in- and out-fields in terms of the local
fields at a finite time are also badly non-local. However, such expressions
are not needed. What is needed are the expressions of the in- and out-fields
in the in- and out-regions, respectively, which are local.

In the present paper, we are going to strengthen the point made in favour of
the Hamiltonian frame based on asymptotic symmetries. We study a field model,
extending thus the cases in which the idea works to infinite-dimensional
systems (see also \cite{prachy}). The simplified model that we investigate
consists of cylindrical gravitational waves, also called sometimes
Einstein-Rosen waves (see, e.g., \cite{beck} and \cite{B}). It seems that this
model can be used as an example of almost everything. Thus, Kucha\v{r} has
studied the embedding variables in canonical theory employing the waves
\cite{KK}. Torre has expressed a complete set of Dirac observables in terms of
local fields \cite{torre}. Ashtekar and Pierri have found a considerable
simplification of Hamiltonian dynamics of the waves using a suitable gauge
\cite{AP}.

Most important, in \cite{AP} an asymptotic boundary term analogous to the ADM
energy has been added to the action for the first time. This is an achievement
because the cylindrical wave spacetimes are not asymptotically flat due to the
cylindrical symmetry so that the well-known results in general relativity
cannot be used. The model can, nevertheless, be reduced from four to three
dimensions by removing the direction of the translational part of the
cylindrical symmetry. The resulting system has the form of gravity coupled to
a scalar field and the geometry of the three-dimensional spacetime is
asymptotically flat in certain sense (locally asymptotically flat). This has
been shown in \cite{AV} and \cite{ABS}, where the boundary term has been
found. Our analysis is, therefore, based on the results of Ashtekar and
Pierri.

In Sec.\ 2, the model of free Newtonian particle moving in one-dimensional
space is used to explain how a symmetry provides a frame for a non-trivial
Hamiltonian dynamics of integrals of motion. The relevant notions and
relations are introduced. The method of reduction by a choice of gauge is also
described within this framework.

Sec.\ 3 summarizes the well-known results on the cylindrical waves that we
shall need, focusing on the asymptotic properties. Sec.\ 4 reviews and
modifies the results of Ashtekar and Pierri so that they become compatible
with the theory of Sec.\ 2. The meaning of {\em asymptotic symmetry} and {\em
  asymptotic time} are explained and their relation to ``ordinary'' symmetry
and time is clarified. The way in which the asymptotic symmetry and time
determine a frame of a Hamiltonian dynamics of Dirac observables for the
cylindrical waves is described. An important new point made in Sec.\ 4 is a
clean separation between gauge transformations and symmetries. In \cite{AP},
\cite{BO'M} and \cite{honnef}, the group of diffeomorphisms has been divided
into gauges and symmetries according to the action of its elements at
infinity. However, it has never been completely clear where the boundary is to
be drawn. When we tried to apply this idea to the cylindrical waves, some
strange paradoxes have appeared. It has turned out that the action must be
parametrized at infinity similarly as in the case of geometrodynamics of
Schwarzschild black holes \cite{KSchw}: privileged spacetime coordinates at
infinity must be added to the set of canonical coordinates of the phase space.
Then, all gauge transformations including reparametrizations at infinity are
generated by constraints, while all asymptotic symmetries are generated by
expressions in the momenta conjugate to the privileged coordinates at
infinity.

The Poisson algebra of the asymptotic Dirac observables listed in Sec.\ 3 is
calculated in Sec.\ 5. The physical phase space is defined by a complete set
of Dirac observables and their Poisson brackets. In Sec.\ 6, the action of the
asymptotic symmetries on the physical phase space is found and the canonical
generator of the continuous group of asymptotic time translations is written
down. The corresponding Hamiltonian dynamics in the physical phase space is
described. It may seem paradoxical that all these gauge-invariant structures
(complete set of Dirac observables, their Poisson brackets and the action of
asymptotic symmetry on them) can be and, indeed, have been calculated from the
gauge-dependent action obtained by Ashtekar and Pierri after a choice of gauge
(cf.\ also \cite{prachy}). The justification thereof is given in Sec.\ 4.

Finally, Sec.\ 7 gives a brief sketch of how the results of Secs.\ 5 and 6 can
be employed for one of possible constructions of quantum theory. The Fock space
and the basic operators on it are specified. The Hamiltonian operator and the
scattering matrix are determined.

\section{Dynamics of Dirac observables \protect\\ in a one-dimensional model}
A finite-dimensional example can illustrate the main aspects of our method.
The underlying geometric framework has been developed for general
finite-dimensional systems in Refs.\ \cite{BO'M}, \cite{JMP1}, \cite{HI} and
\cite{DDO}.

Consider a free Newtonian particle of unit mass in one-dimensional space
evolving in the Newtonian time $T$. Let its position be denoted by $Q$.
The system is described by the
action
\begin{equation}
S[Q] =  \frac{1}{2} \, \int_{T_1}^{T_2} \, dT \,  \big( Q_{,T} \big)^2  \, .
\label{priv action}
\end{equation}
This action is invariant under the translation
\begin{equation}
T \rightarrow T + \tau
\label{priv sym}
\end{equation}
for any time $T$ and any real parameter $\tau$. This is a one-dimensional
group of symmetry that will play a crucial role in what follows. The symmetry
implies via Noether theorem the conservation of energy
$$
  E = \frac{1}{2}Q^2_{,T}\ .
$$
The corresponding Hamiltonian action is
\begin{equation}
S[Q, P_{Q}] = \int_{T_1}^{T_2} \, dT \, \Big( \, P_{Q} \, Q_{,T} -
\frac{1}{2} \, {P}_{Q}^2 \Big) \, .
\label{reduced action}
\end{equation}
The one-form $\Theta = P_Q\,dQ$ contained in the action is called the
Liouville form.  The space with the coordinates $Q$ and $P_Q$ is the {\em
  physical phase space} $\Gamma_1$; it carries the symplectic two-form
$\Omega_1 = d\Theta = dP_Q\wedge dQ$.

The equations of motion implied by action (\ref{reduced action}) are
\begin{eqnarray}
Q_{,T} &=& P_{Q}\ ,
\label{T evolution1}
\\
P_{Q,T} &=& 0\ .
\label{T evolution2}
\end{eqnarray}
Their general solution is
\begin{eqnarray}
Q(T) &=& q + p  \, T\ ,
\label{T solution1}
\\
P_{Q}(T) &=& p \ ,
\label{T solution2}
\end{eqnarray}
where $(q,p) \in {{\mathbb R}}^2$ are constant for a particular solution. We
can, therefore, define the {\em space of solutions} $\Gamma_2$ as ${{\mathbb
    R}}^2$ with coordinates $q$ and $p$. An important observation is that
$\Gamma_2$ also carries a symplectic structure.  Indeed, Eqs.\ (\ref{T
  solution1}) and (\ref{T solution2}) can be considered as a $T$-dependent
canonical transformation because they imply
$$
  P_Q\,dQ = p\,dq + d\,\frac{(Q-q)^2}{2T}\ ,
$$
where, of course, the action of the differential $d$ ``hits'' only the
variables $Q$ and $q$, not $T$; $(Q-q)^2/2T$ is the generating function. The
resulting symplectic form $\Omega_2$ of $\Gamma_2$ is $dp\wedge dq$. The
dynamical change $\delta_d$ within the time increment $\delta T$ in $\Gamma_2$
is trivial:
\begin{equation}\label{3,1}
  \delta_d p = 0\ ,\quad \delta_d q = 0\ .
\end{equation}

We can also consider the solutions (\ref{T solution1}) and (\ref{T solution2})
as defining two scalar fields on the one-dimensional time manifold ${\mathbb
  R}$ with coordinate $T$, the so-called {\em background manifold}. The push
forward by the symmetry transformation (\ref{priv sym}) acts on these fields
as follows
\begin{eqnarray}
Q(T) &\mapsto& Q(T-\tau)\ ,
\label{4,1}
\\
P_{Q}(T) &\mapsto& P_{Q}(T-\tau) \ .
\label{4,2}
\end{eqnarray}
This symmetry action preserves, of course, the property of being a solution of
the equations of motion; if $q$ and $p$ describe the untransformed solution
and $q'$ and $p'$ the transformed one, then Eqs.\ (\ref{T solution1}) and
(\ref{T solution2}) yield
\begin{equation}\label{4,3}
  q' = q - p\tau\ ,\quad p' = p\ .
\end{equation}
This action of the symmetry transformation on the space $\Gamma_2$ is
canonical and is generated by the function $-p^2/2$ (as the Poisson brackets
show):
\begin{eqnarray}\label{4,4}
\delta_s q &=& -p\delta\tau = \{q,-\frac{\delta\tau}{2}p^2\}\ ,
\\ \label{4,5}
\delta_s p &=& 0 = \{p,-\frac{\delta\tau}{2}p^2\} \ ,
\end{eqnarray}
where $\delta_s$ is the change caused by the symmetry transformation.

The key observation is now the following. If we compare the time change in $q$
and $p$ due to the dynamics (Eq.\ (\ref{3,1})) with that due to the symmetry
((Eqs.\ (\ref{4,4}) and (\ref{4,5})), we find that the relative time changes
$\delta_d q - \delta_s q$ and $\delta_d p - \delta_s p$ are formally identical
to the original dynamics generated by the Hamiltonian $P_Q^2/2$ in $\Gamma_1$.

The description of the dynamics of our model can be made generally covariant
by {\em parametrization} (see, e.g., \cite{banf}). An arbitrary time $t =
t(T)$ is introduced and the action (\ref{priv action}) is brought into the
{\em constraint-Hamiltonian form} (i.e., the Hamiltonian is a linear
combination of constraint functions---a typical property of generally
covariant systems, see \cite{HT}):
\begin{equation}\label{param action}
\begin{split}
&S[Q, P_Q, T, P_T; N] =\\
&\qquad= \int_{t_1}^{t_2} \, dt \, \Big( \, P_Q \,
Q_{,t} + P_T \, T_{,t} - N \, \big( P_T + \frac{1}{2} \,
{P_Q^2}  \big) \, \Big) \, .
\end{split}\raisetag{40pt}
\end{equation}
Let the space $\cal P$ consist of all initial data $Q$, $P_Q$, $T$ and $P_T$
for the field equations implied by action (\ref{param action}). The space
$\cal P$ equipped with the symplectic form ${\Omega}_{\cal P}$ derived from
the Liouville form of action (\ref{param action}),
$$
  {\Omega}_{\cal P} = dP_Q \wedge dQ + dP_T \wedge dT\ ,
$$
is called the {\em extended phase space} of the system. $N$ is a Lagrange
multiplier and $C = 0$ is a constraint with
\begin{equation}
  C = P_T + \frac{1}{2} \, P_Q^2
\label{initial constraint}
\end{equation}
being the constraint function. The surface ${\cal C}$ in ${\cal P}$ defined by
the constraint is called the {\em constraint surface}. Since equation $C = 0$
can be solved for $P_T$, we can choose the functions $Q$, $P_Q$ and $T$ as
coordinates on ${\cal C}$.

The transformations that are canonically generated by the function ${\mathcal
  H}(N) = NC$ can be considered as reparametrizations---general changes of the
time parameter. They are, therefore, analogous to {\em gauge transformations}.
The corresponding gauge group acts along the constraint surface ${\cal C}$. At
the same time, its {\em orbits} in ${\cal C}$ coincide with the solutions
(\ref{T solution1}) and (\ref{T solution2}). The physical phase space
$\Gamma_2$ of gauge-equivalent solutions can, therefore, be identified with
the quotient of the constraint surface by the gauge orbits:
\begin{equation}
\Gamma_2 = \frac{{\cal C}}{{\rm Orb}} \ .
\label{phys space2}
\end{equation}
In coordinates $(q,p)$ on $\Gamma_2$ and $(Q,P_Q,T)$ on $\cal C$, the
projection ${\rm Proj}_{({\cal C} \, \rightarrow \, \Gamma_2)}$ derived from
(\ref{phys space2}) is
\begin{equation}
\Big( Q \, , \, P_Q \, , \, T \Big) \in {\cal C} \, \mapsto \, \Big( q \, ,
\, p \Big) = \Big( Q - P_Q \, T \, , \, P_Q \Big) \in \Gamma_2 \ .
\label{Projection}
\end{equation}
The symplectic form $\Omega_2$ on $\Gamma_2$ can be obtained from
$\Omega_{\mathcal P}$ as follows. First, ${\Omega}_{\cal P}$ is pulled back
from ${\cal P}$ to $\cal C$. This yields a two-form $\Omega_{\cal C}$
degenerated along the gauge orbits in $\cal C$.  The projection ${\rm
  Proj}_{({\cal C} \, \rightarrow \, \Gamma_2)}$ in (\ref{Projection})
determines the symplectic form ${\Omega}_2$ as the unique solution of the
equation ${\Omega}_{\cal C} = {\rm Proj}^{*}_{({\cal C} \, \rightarrow \,
  \Gamma_2)}\, {\Omega}_2$ where $^*$ denotes the pull-back mapping.

We shall also need the concept of {\em transversal surface} ${\mathcal T}
\subset {\mathcal C}$. This is a section ${\sigma}_{(\Gamma_2 \, \rightarrow
  \, {\cal C})} : \Gamma_2 \mapsto {\mathcal C}$ in the sense that
\begin{equation}
{\rm Proj}_{({\cal C} \, \rightarrow \, \Gamma_2)} \circ {\sigma}_{(\Gamma_2 \,
  \rightarrow \, {\cal C})} = {\rm Id}_{\Gamma_2}
\label{transversal section}
\end{equation}
with respect to the projection (\ref{Projection}). Whenever $\cal C$ admits
such a section, the surface $\mathcal T = {\sigma}_{(\Gamma_2 \, \rightarrow
  \, {\cal C})}(\Gamma_2)$ is a copy of the physical phase space $\Gamma_2$.
A bijection between any surface $\mathcal T$ and the physical phase space
$\Gamma_2$ can be defined by restricting the projection ${\rm Proj}_{({\cal C}
  \, \rightarrow \, \Gamma_2)}$ to each particular $\mathcal T$. The
symplectic form $\Omega_2$ induces through each such bijection a unique
symplectic form $\Omega_{\mathcal T}$. In this way each $\mathcal T$ also
becomes a phase space with a symplectic structure that is isomorphic to that
of $\Gamma_2$. Transversal surface $\mathcal T$ is called {\em regular} if if
it is not tangential to gauge orbits at any of its points. If $\mathcal T$ is
defined by the equation $F(Q,P_Q,T) = 0$, then the regularity condition is a
non-vanishing Poison bracket
\begin{equation}
  \{F,{\mathcal H}(N)\}|_{\mathcal C} \neq 0\quad \forall N\ .
\label{**}
\end{equation}
The meaning of the regularity of transversal surfaces simply is that the gauge
condition breaks the gauge completely.  Systems that are not pathological
possess many transversal surfaces.

One way to quantize a generally covariant system is to reduce it to an
unconstrained system of the kind (\ref{reduced action}). We shall now show two
methods of reduction: by a gauge condition and via Dirac observables and
symmetry.

\subsection{Reduction by a choice of gauge}
A {\em gauge condition} is a choice of a particular family of regular
transversal surfaces that foliate $\mathcal C$. Let the family be given by
the set of equations
\begin{equation}
  \tilde{F}(Q,P_Q,T) = \tilde{T} \ ;
\label{8,1}
\end{equation}
for each fixed real $\tilde{T}$, one surface ${\mathcal T}_{\tilde{T}}$ of the
family is defined. The reduction using the condition (\ref{8,1}) can proceed
as follows. Suppose that a canonical transformation in $\mathcal P$ is known
between the original variables $(Q, P_Q, T, P_T)$ and some canonical
coordinates $(\tilde{q}, \tilde{p}, \tilde{T}, \tilde{P})$ that have been
chosen so as to contain $\tilde{T}$. Then, using the regularity condition of
the gauge, one can show that the constraint $C = 0$ is solvable with respect
to $\tilde{P}$ and so can equivalently be written as
$$
  \tilde{P} + \tilde{\mathcal H}(\tilde{T},\tilde{q},\tilde{p}) = 0\ ,
$$
where $\tilde{\mathcal H}$ is some smooth function. The canonical
transformation brings action (\ref{reduced action}) to the form
$$
  S(\tilde{q},\tilde{p},\tilde{T},\tilde{P},\tilde{N}) =
  \int_{t_1}^{t_2}dt\,[\tilde{p}\tilde{q}_{,t} + \tilde{P}\tilde{T}_{,t}-
  \tilde{N}(\tilde{P}+\tilde{\mathcal H})]\ ,
$$
where $\tilde{N}$ is a new Lagrange multiplier defined by
$$
  NC = \tilde{N}(\tilde{P}+\tilde{\mathcal H})\ .
$$
Next, $\tilde{T}$ is chosen as the integration variable $t$ and the
action is restricted to the constraint surface. The result is
\begin{equation}
  \tilde{S}(\tilde{q},\tilde{p}) =
  \int_{\tilde{T}_1}^{\tilde{T}_2}d\tilde{T}\,(\tilde{p}\tilde{q}_{,\tilde{T}}
  -\tilde{\mathcal H})\ .
\label{9,3}
\end{equation}
By this, the reduction is finished.

A problem with this kind of reduction is that the new variables $\tilde{q}$,
$\tilde{p}$ as well as the new time $\tilde{T}$ are not the same as the
original variables $Q$, $P_Q$ and $T$. Classically, the two actions
(\ref{9,3}) and (\ref{reduced action}) are equivalent, because they are
related by an extended gauge transformation. The two quantum mechanics,
however, that are obtained by the standard quantization method from them, {\em
  cannot} be unitarily equivalent \cite{paris}: the transformation (\ref{8,1})
between the respective times involves operators, while each of the times must
be a parameter in the respective quantum mechanics.

\subsection{Reduction using Dirac observables and symmetry}
A {\em Dirac observable} $o({\cal P})$ is a function $o:{\cal P} \mapsto
{\mathbb R}$ whose Poisson bracket with the constraint $C$ vanishes when
restricted to $\cal C$. Dirac observables are gauge invariant.

The correspondence between Dirac observables on $\cal P$ and functions on
$\Gamma_2$ is the following: Each function $f : \Gamma_2 \mapsto {\mathbb
  R}$ determines a function $f\circ$ ${\rm Proj}_{({\cal C} \, \rightarrow \,
  \Gamma_2)}$ on the constraint surface via the projection mapping
(\ref{Projection}). In the chart $(Q,T,P_Q)$ on $\cal C$ such a function has
the form
\begin{equation}
  f(Q-P_Q T,P_Q) \ .
\label{complete C}
\end{equation}
The next step is to extend this function from $\mathcal C$ to $\mathcal P$.
Let us denote such an extension by $o : {\mathcal P} \mapsto {\mathbb R}$. The
only condition is that $o$ be a smooth function on $\mathcal P$ the values of
which coincide with $f(Q-P_Q T,P_Q)$ at $\mathcal C$. In this way, the
original function $f$ on $\Gamma_2$ defines an equivalence class $\{o\}$ of
functions $o$ on $\mathcal P$. One can show that any two elements $o_1$ and
$o_2$ of $\{o\}$ satisfy $o_2 = o_1 + NC$, where $N$ is a smooth function on
$\mathcal P$ (see, e.g., \cite{HT}). It also follows immediately from the
construction that $o$ is constant along orbits so that it is a Dirac
observable. Thus, a whole class of Dirac observables corresponds to one
function on $\Gamma_2$ (one often speaks about Dirac observables meaning these
classes).

On the other hand, each Dirac observable $o$ defines via the restriction to
$\cal C$ a function on $\cal C$ that is constant along orbits. Such a function
determines, in turn, a unique function $f$ on $\Gamma_2$.

The Poisson brackets between Dirac observables can be calculated using the
symplectic structure of the extended phase space $\mathcal P$. It is easy to
show \cite{JMP1} that the Poisson bracket $\{o_1,o_2\}$ of two Dirac
observables is again a Dirac observable and that the Poisson brackets
$\{o_1,o_2\}$ and $\{o_1 + N_1C,o_2 + N_2C\}$ lie in the same class. Thus, the
Poisson brackets between the equivalence classes $\{o\}$ are well defined.
Moreover, if the class with the representative $o_i$ corresponds to the
function $f_i$, $i=1,2$, on $\Gamma_2$, and the class with the representative
$\{o_1,o_2\}$ corresponds to $f$, then
$$
  \{f_1,f_2\}_{\Gamma_2} = f\ ,
$$
as is shown in Ref.\ \cite{JMP1}.  It follows from this that a {\em complete
  set of Dirac observables}, together with their Poisson algebra, determine the
structure of the physical phase space $\Gamma_2$. A complete set of Dirac
observables separates gauge orbits, and in simple cases can be used as a
coordinate system on the quotient space ${\mathcal C}/\text{Orb} = \Gamma_2$.
For our model, a complete set is formed by the functions $o_0 = Q-P_QT + N_0C$
and $o_1 = P_Q + N_1C$, where $N_0$ and $N_1$ are smooth functions on
$\mathcal P$. A simple calculation gives that the only non-trivial bracket is
$$
  \{o_0,o_1\} = 1 + NC\ ,
$$
where $N = \{N_0,o_1\} + \{o_0,N_1\}$; the Dirac observables $o_0$ and
$o_1$ correspond to the functions $q$ and $p$ on $\Gamma_2$.

The symmetry group (\ref{priv sym}) acts on the extended phase space as
follows
\begin{equation}\label{*}
  (Q,P_Q,T,P_T) \mapsto (Q,P_Q,T+\tau,P_T)\ ,
\end{equation}
and is, therefore, generated by the momentum $P_T$ conjugate to $T$. Observe
that $P_T$ itself is a Dirac observable; one can prove \cite{JMP1} that any
continuous symmetry group is generated by a Dirac observable. Now, $P_T$ has a
non-trivial action on Dirac observables. For example,
\begin{eqnarray*}
  \{o_0,P_T\} &=& -o_1 + \bar{N}_0C\ , \\
  \{o_1,P_T\} &=&  \bar{N}_1C\ ,
\end{eqnarray*}
where $\bar{N}_0$ and $\bar{N}_1$ are suitable functions on $\mathcal P$. The
change of Dirac observables {\em referred to} the symmetry as ``zero motion''
is, therefore, non-trivial. It is easy to see that this change is generated by
the function $-P_T$. The value of $P_T$ at $\mathcal C$ is, however,
$$
  -P_T|_{\mathcal C} = \frac{1}{2}P_Q^2\ ,
$$
and it lies in the class $o^2_1/2 + NC$.  The corresponding function on
$\Gamma_2$ is, therefore, $H = p^2/2$, and it plays the role of the
Hamiltonian of the constructed dynamics. In this way, we have recovered the
dynamics and the phase space $\Gamma_1$ of the original system so that the
reduction is accomplished.

Mathematically, any symmetry of ${\cal P}$ that is not pure gauge
transformation can generate a non-trivial evolution of Dirac observables on
$\Gamma_2$ because it defines a non-trivial mapping between gauge orbits in
${\cal C}$. By projection to $\Gamma_2$ a symmetry is obtained which can be
interpreted as the generator of a dynamical evolution on $\Gamma_2$.
Physically, it must be additionally required that the symmetry be privileged
by the situation at hand. Only then its role as a true Hamiltonian on
$\Gamma_2$ can be justified.  Here, the constant translation (\ref{*}) is
physically privileged by the arguments leading from (\ref{priv action}) to
(\ref{param action}) and in particular by the fact that it yields through
Noether's theorem the energy of the Newtonian particle in the privileged
reference system $T$. The transformation (\ref{*}) is indeed a symmetry of
$\cal P$ because the Poisson bracket of its generator, $-P_T$, with the
constraint function (\ref{initial constraint}) vanishes on $\cal C$.

The following observation is very important. If $P_T$ generates a symmetry
that leads to the Hamiltonian $H$ in $\Gamma_2$, then so does $P_T + N'C$ for
any smooth $N'$: the dynamics of Dirac observables is uniquely determined by
the whole class of symmetry generators. Why is this important: In our simple
model, we have a unique symmetry and it is generated by $P_T$. The reason is
that our model is a so-called ``already parametrized system'' with a
privileged time $T$. Indeed, there also is a privileged choice of gauge due to
this fact: $F(Q,P_Q,T) \equiv T$, which leads to the ``right'' action
(\ref{reduced action}) by the reduction procedure of Sec.\ 2.1. However, many
models of real interest, such as general relativity, are not already
parametrized systems \cite{HKij}. For such models, there is no privileged time
and no symmetry in general (cf.\ \cite{nosym}). But in asymptotically flat
cases, there is a privileged asymptotic time and an asymptotic symmetry. As it
is shown in \cite{BO'M}, such symmetries do not determine their generators in
the extended phase space $\mathcal P$ uniquely but only up to addition of a
linear combination of constraints. Despite that, they still define a unique
dynamics of Dirac observables.

\section{Polarized cylindrical waves:  solutions and asymptotic behavior}
A vacuum spacetime describing cylindrical gravitational waves with a fixed
state of polarization (one degree of freedom per point) has two commuting,
hypersurface-orthogonal spacelike Killing vectors $\partial/\partial\varphi$
and $\partial/\partial z$; the Killing field $\partial/\partial\varphi$ is
rotational and it keeps a timelike axis fixed; $\partial/\partial z$ is
translational; coordinates $\varphi$ and $z$ are invariantly defined up to
a translation $z \mapsto z + a$. The metric can be written in the form
\begin{equation}\label{3.1}
  ds^2 = e^{\gamma - \psi}(-dT^2 + dR^2) + e^\psi dz^2 + R^2
  e^{-\psi}d\varphi^2 \ ,
\end{equation}
where $T$ and $R$ are invariantly defined, $T$ up to a translation $T \mapsto
T + a$. In the above equation, $\psi = \psi(T,R)$ and $\gamma = \gamma(T,R)$.
(To obtain Eq.\ (\ref{3.1}) one uses a consequence of vacuum field
equations---see, e.g., \cite{KK}, \cite{BIS}.)

It is well-known that because of the translational symmetry $\partial/\partial
z$, the four-dimensional Einstein equations are equivalent to the
three-dimensional Einstein equations with certain matter sources (see, e.g.,
\cite{ABS}, \cite{AP} and \cite{B}). In our case of cylindrical symmetry
($\partial/\partial\varphi$ is a further Killing field) the four-dimensional
Einstein vacuum equations the solutions of which give Einstein--Rosen waves
are equivalent to Einstein equations in three dimensions with a zero-rest-mass
scalar field $\psi$ as a source. It is, however, more advantageous for the
canonical formulation to work with the physical Klein-Gordon field $\phi =
\psi/\sqrt{8G}$, $G$ being the Newton constant. Hence, we formulate
everything with the help of the field $\phi$.

In three dimensions, the metric is given by (cf.\ \cite{ABS} and \cite{B})
\begin{equation}\label{3.2}
  ds^2 = g_{ab}dx^adx^b = e^\gamma(-dT^2 + dR^2) + R^2d\varphi^2\ ,
\end{equation}
and the Einstein field equations become
\begin{eqnarray}\label{3.3}
  \frac{\partial^2\gamma}{\partial R^2} - \frac{\partial^2\gamma}{\partial
  T^2} +  \frac{1}{R}\frac{\partial\gamma}{\partial R} & = &
  8G\left(\frac{\partial\phi}{\partial T}\right)^2\ , \\
\label{3.4}
  -\frac{\partial^2\gamma}{\partial R^2} + \frac{\partial^2\gamma}{\partial
  T^2} +  \frac{1}{R}\frac{\partial\gamma}{\partial R} &=&
  8G\left(\frac{\partial\phi}{\partial R}\right)^2\ , \\
\label{3.5}
  \frac{1}{R}\frac{\partial\gamma}{\partial T} &=&
  8G\frac{\partial\phi}{\partial R}\frac{\partial\phi}{\partial T}\ .
\end{eqnarray}
The field equation for $\phi$,
\begin{equation}\label{3.6}
  -\frac{\partial^2\phi}{\partial T^2} + \frac{\partial^2\phi}{\partial R^2} +
  \frac{1}{R}\frac{\partial\phi}{\partial R} = 0\ ,
\end{equation}
is the wave equations for the nonflat metric (\ref{3.2}) {\it as well as} for
the flat (Minkowski) metric obtained by putting $\gamma = 0$ in
(\ref{3.2}). This crucial simplification implies that the scalar field $\phi$
is decoupled from the equations satisfied by the metric. Eqs.\
(\ref{3.3})--(\ref{3.5}) reduce to two simple equations
\begin{eqnarray}\label{3.7}
  \frac{\partial\gamma}{\partial R} &=& 4G
  R\left[\left(\frac{\partial\phi}{\partial T}\right)^2 +
  \left(\frac{\partial\phi}{\partial R}\right)^2\right]\ , \\
\label{3.8}
  \frac{\partial\gamma}{\partial T} &=& 8GR
  \frac{\partial\phi}{\partial T}
  \frac{\partial\phi}{\partial R} \ ,
\end{eqnarray}
the wave equation (\ref{3.6}) is their integrability condition. We can thus
solve the axisym\-metric---in three dimensions `spherically' symmetric---wave
equation (\ref{3.6}) on Minkow\-ski space and then solve Eqs.\ (\ref{3.6}) and
(\ref{3.8}) for the metric function $\gamma(T,R)$ by quadratures. These
well-known facts are of key importance in the canonical and quantum theory
since all physical degrees of freedom are contained in the scalar field.

We shall now briefly review some of the results on the asymptotics obtained in
\cite{ABS} and \cite{ABSI}. We shall extend the discussion by including both
future and past null infinities, and later also by employing a Fourier-type
decomposition.

The Cauchy data for the scalar field $\phi$, given on the Cauchy surface
topologically ${{\mathbb R}}^2$, suffice to determine the whole spacetime
metric. For data which fall off appropriately, the three-dimensional
Lorentzian geometry is asymptotically flat both at spatial \cite{AV} and null
infinity \cite{ABS} although in four dimensions the Einstein-Rosen spacetimes
are not asymptotically flat (see \cite{ABSI} for a detailed investigation of
cylindrical waves at null infinity in four dimensions).

By employing the ``method of descent'' from the Kirchhoff formula in four
dimensions one can find the representation of the solution $\phi(T,R)$ of the
wave equation (\ref{3.6}) in three dimensions in terms of Cauchy data $\phi_0
=\phi(0,R)$ and $\phi_1=\phi_{,T}(0,R)$. This has been used in \cite{ABS} to
find the asymptotic behavior of the field $\phi$ and the whole metric
(\ref{3.2}) at the future null infinity for the data of compact support (see
Sec. II in \cite{ABS}). By applying the same procedure one can analyze the
solutions at the past null infinity. Introducing retarded and advanced time
coordinates
\begin{equation}\label{3.9}
  U = T - R\ ,\quad V = T + R
\end{equation}
(notice that these are null coordinates for both flat Minkowski metric and the
curved metric (\ref{3.2})), one obtains expansions in the powers of $R^{-1/2}$
along null hypersurfaces $U=$ constant and $V=$ constant of the form
\begin{eqnarray}\label{3.10}
  \phi(V,R) &=& \frac{1}{\sqrt{R}}\ g(V) + \sum_{k=1}^\infty
  \frac{g_k(V)}{R^{k+1/2}}\ , \\
\label{3.11}
  \phi(U,R) &=& \frac{1}{\sqrt{R}}\ f(U) + \sum_{k=1}^\infty
  \frac{f_k(U)}{R^{k+1/2}}\ .
\end{eqnarray}
The coefficients in the expansions are determined by the Cauchy data. By
rewriting the Einstein field equations (\ref{3.7}) and (\ref{3.8}) in terms of
$U$ and $R$ (respectively $V$ and $R$), we obtain the asymptotic behaviour of
the metric function $\gamma$ at ${\mathcal I}^+$ in the form
\begin{equation}\label{3.12}
  \gamma(U,R) = \gamma_\infty - 8G\int_{-\infty}^U
  dU\,\left(\frac{df}{dU}\right)^2 + O(R^{-2})\ ,
\end{equation}
and similarly at ${\mathcal I}^-$. Here the constant $\gamma_\infty$, which
will play a key role in the following, is determined uniquely by the Cauchy
data for $\phi$ (cf. Eq.\ (\ref{3.7}))
\begin{equation}\label{3.13}
  \gamma_\infty = \gamma(0,\infty) = 4G\int_0^\infty
  dR\,R\left[\left(\frac{\partial\phi}{\partial T}\right)^2 +
  \left(\frac{\partial\phi}{\partial R}\right)^2\right]\ .
\end{equation}

The value of $\gamma_\infty$ represents the total energy of the scalar field
$\phi$ computed by using the Minkowski metric. For any nontrivial data,
$\gamma_\infty$ is positive. Hence, the metric at spatial infinity, given by
\begin{equation}\label{3.14}
  ds^2 = e^{\gamma_\infty}(-dT^2 + dR^2) + R^2 d\varphi^2 \ ,
\end{equation}
has a conical singularity because the distance of the circles with radii $R$
and $R+dR$ is different by a factor $e^{\gamma_\infty}$ from the difference
of their circumferences divided by $2\pi$. It can be shown \cite{ABS} that as
one approaches ${\mathcal I}^+$ ($R \rightarrow \infty$, $U=$ constant), one
finds (cf.\ Eq.\ (\ref{3.12}))
\begin{equation}\label{3.15}
  \gamma(U,\infty) = \gamma_\infty - 8G\int_{-\infty}^U
  du\,\left(\frac{df}{du}\right)^2\ ,
\end{equation}
and $\gamma$ to vanish at the timelike infinity $i^+$. Hence,
\begin{equation}\label{3.16}
  \gamma_\infty = 8G\int_{-\infty}^\infty
  dU\,\left(\frac{df}{dU}\right)^2\ .
\end{equation}
The conical singularity, present at spacelike infinity, is thus ``radiated
out'', and the future timelike infinity $i^+$ becomes smooth. Eq.\
(\ref{3.15}) plays the role of the well-known Bondi mass-loss formula, the
function $df/dU$ being analogous to the Bondi news function (see also
\cite{St}, Eq.\ (3.6), for an analysis in four dimensions). Clearly, analogous
formulae to (\ref{3.15}) and (\ref{3.16}) are valid for incoming waves, with
$df/dU$ replaced by $dg/dV$:
$$
  \gamma(V,\infty) = 8G\int_{-\infty}^V
  dv\,\left(\frac{dg}{dv}\right)^2\ ,
$$
and
\begin{equation}\label{3.17}
  \gamma_\infty = 8G\int_{-\infty}^\infty
  dV\,\left(\frac{dg}{dV}\right)^2\ .
\end{equation}
Here we assume smooth past timelike infinity $i^-$ and incoming waves from the
past null infinity ${\mathcal I}^-$ with a null data $g(V)$ bring in
mass-energy which reveals itself as a conical singularity characterized by
$\gamma(V,\infty)$. At spatial infinity $i_0$ ($V=\infty$, $R=\infty$) this
becomes just the constant $\gamma_\infty$ given in Eq.\ (\ref{3.13}). The
fluxes of radiation, the analogues of the news function, as well as conical
singularities are observable quantities at the past and future null
infinities. Both are given by the asymptotic null data $g(V)$ and $f(U)$. The
asymptotic null data will be important in the following.

Starting from the representation of the solutions of the three-dimensional
wave equation (\ref{3.6}) in terms of the Kirchhoff-type formula obtained by
the ``method of descent'' from four dimensions one can, for the Cauchy data of
compact support, obtain not only expansions (\ref{3.10}) and (\ref{3.11}), but
also the explicit expression for the null data $f(U)$ (respectively $g(V)$) as
the integral over the Cauchy data $\phi_0$ and $\phi_1$. However, these
integrals become simple only for retarded times $U$ so large that the support
of the data is completely in the interior of the past cone (similarly for the
advanced times at ${\mathcal I}^-$), see \cite{ABS}. Here we need the null data
for all $U$'s at ${\mathcal I}^+$ and $V$'s at ${\mathcal I}^-$.

To achieve this, we start from a Fourier-type decomposition. This, in three
dimensions, means to write the solutions in terms of the Bessel functions of
zero order provided that we require the solutions to be regular everywhere, in
particular at $R=0$ (see, e.g., \cite{J}).

Thus, we start from the solutions of the form
\begin{equation}\label{3.18}
  \phi(T,R) = \frac{1}{\sqrt{2}}\int_0^\infty
  d\omega\,\left[A(\omega)J_0(\omega R)e^{-i\omega T} +\cxconj\right]\ ;
\end{equation}
as usual, we write just ``$\cxconj$'' instead of the second term, meaning the
complex conjugate of the first one. Using the asymptotic expansion of the
Bessel function at $R \rightarrow \infty$ (see, e.g., \cite{J}), we obtain
\begin{equation}\label{3.19}
\begin{split}
  &\phi(T,R) = \frac{1}{2\sqrt{R\pi}}\int_0^\infty
  \frac{d\omega}{\sqrt{\omega}}\,
  \biggl\{\left[A(\omega)e^{-i(\pi/4) -i\omega U} + \cxconj\right] \\
  &\qquad+ \left[A(\omega)e^{i(\pi/4) -i\omega V} + \cxconj\right]\biggr\} +
  O(R^{-3/2}) \ ,
\end{split}\raisetag{20pt}
\end{equation}
where $U$ and $V$ are retarded and advanced time coordinates given by Eq.\
(\ref{3.9}). Hence, the null data at the future and past null infinities read
as follows:
\begin{align}\label{3.20}
  f(U) &= \frac{1}{2\sqrt{\pi}}\int_0^\infty
  \frac{d\omega}{\sqrt{\omega}}\,\left[A(\omega)e^{-i(\pi/4) -i\omega U} +
  \cxconj\right] \ , \\
\label{3.21}
  g(V) &= \frac{1}{2\sqrt{\pi}}\int_0^\infty
  \frac{d\omega}{\sqrt{\omega}}\,\left[A(\omega)e^{i(\pi/4) -i\omega V} +
  \cxconj\right] \ .
\end{align}
It is easy to invert the last equations by writing, for example,
\begin{align}\label{3.22}
  g(V) + g(-V) &= \sqrt{\frac{2}{\pi}}\int_0^\infty
  d\omega\,[\tilde{A}(\omega) +\tilde{A}^*(\omega)]\cos\omega V\ , \\
\label{3.23}
  g(V) - g(-V) &= -i\sqrt{\frac{2}{\pi}}\int_0^\infty
  d\omega\,[\tilde{A}(\omega) -\tilde{A}^*(\omega)]\sin\omega V\ ,
\end{align}
where $\tilde{A}(\omega) = (2\omega)^{-1/2}A(\omega)e^{i\pi/4}$. Using Fourier
cosine and sine (inverse) transforms to express $\tilde{A} \pm \tilde{A}^*$,
we find
\begin{equation}\label{3.24}
  A(\omega) = \sqrt{\frac{\omega}{\pi}}e^{-i\pi/4}\int_0^\infty
  dV\left[g(V)e^{i\omega V} + g(-V)e^{-i\omega V}\right]\ ,
\end{equation}
$A^*(\omega)$ being given by complex conjugation. Alternatively, we can write
\begin{equation}\label{3.25}
  A(\omega) = \sqrt{\frac{\omega}{\pi}}e^{-i\pi/4}\int_{-\infty}^\infty
  dV\,g(V)e^{i\omega V}\ .
\end{equation}
Similarly, from Eq.\ (\ref{3.20}) we get
\begin{equation}\label{3.26}
  A(\omega) = \sqrt{\frac{\omega}{\pi}}e^{i\pi/4}\int_0^\infty
  dU\left[f(U)e^{i\omega U} + f(-U)e^{-i\omega U}\right]\ ,
\end{equation}
or
\begin{equation}\label{3.27}
  A(\omega) = \sqrt{\frac{\omega}{\pi}}e^{i\pi/4}\int_{-\infty}^\infty
  dU\,f(U)e^{i\omega U}\ .
\end{equation}
Since according to Eq.\ (\ref{3.18}) the functions $A(\omega)$ determine
the solutions $\phi(T,R)$ everywhere, Eqs.\ (\ref{3.24})--(\ref{3.27}) imply
that either the null data $g(V)$ at ${\mathcal I}^-$ or $f(U)$ at ${\mathcal
  I}^+$ determine $\phi(T,R)$ uniquely in the spacetime.

The amplitudes $A(\omega)$ can be expressed also in terms of the Cauchy data
$\phi_0 = \phi(0,R)$ and $\phi_1 = \phi_{,T}(0,R)$ directly from Eq.\
(\ref{3.18}). Using the Hankel transform (see, e.g., \cite{J}): for two
functions $X(x)$ and $Y(y)$,
\begin{equation}\label{HT1}
X(x) = \int_0^\infty dy\,Y(y)\sqrt{xy}J_0(xy)
\end{equation}
is equivalent to
\begin{equation}\label{HT2}
Y(y) = \int_0^\infty dx\,X(x)\sqrt{xy}J_0(xy)\ .
\end{equation}
Expressing $\phi_1$ from Eq.\ (\ref{3.18}) we obtain
\begin{equation}\label{3.28}
  A(\omega) = \frac{1}{\sqrt{2}}\int_0^\infty dR\,(\omega\phi_0 -
  i\phi_1)RJ_0(\omega R)\ .
\end{equation}
Hence, as expected, we need both $\phi_0$ and $\phi_1$ to determine the
solution of the wave equation (\ref{3.6}) everywhere. For time-symmetric
initial data, $\phi_1 = 0$, the amplitudes $A(\omega)$ become real.

Although for the Cauchy data of compact support and even for more general data
falling off sufficiently rapidly at spatial infinity we get $\phi \sim
1/\sqrt{R}$ at null infinities as in Eqs.\ (\ref{3.10}) and (\ref{3.11}) (see
\cite{ABS}), at spatial infinity, i.e.\ in the limit $R \rightarrow \infty$,
$T$ fixed, the solutions fall off more rapidly:
\begin{equation}\label{3.29}
  \phi \sim O(1/{R}),\quad \phi_{,R} \sim O(1/{R^2})\ .
\end{equation}
This will be needed in the canonical theory. To demonstrate the fall-off,
employ again the asymptotic form of $J_0(\omega R)$ at $R \rightarrow \infty$
in Eq.\ (\ref{3.18}),
$$
  \phi =
  \frac{\text{const}}{\sqrt{R}}\int_0^\infty\frac{d\omega}{\sqrt{\omega}}
  \cos\bigl(\omega R - \frac\pi4\bigr)\bigl[A(\omega)e^{-i\omega T} + \cxconj\bigr]\ ,
$$
and put $\omega R = \omega'$ in the integral. Then we get
$$
  \phi =\!
  \frac{\text{const}}{\sqrt{R}}\!\int_0^\infty\!\!\!\frac{d\omega'}{\sqrt{R\omega'}}
  \cos\bigl(\omega' \!-\! \frac\pi4\bigr)
  \bigl[A({\textstyle\frac{\omega'}{R}})e^{-i\omega' {T}/{R}} + \cxconj\bigr]\,,
$$
which at large $R$ and fixed $T$ leads to $\phi \sim 1/R$.

Finally, let us illustrate previous results by a simple example. The
Weber-Wheeler-Bonnor pulse \cite{WW}, \cite{Bo} represents an exact,
time-symmetric vacuum solution of the Einstein equations with cylindrical
symmetry which satisfies all regularity conditions required above. The pulse
comes in from the past null infinity, concentrates around the axis of symmetry
(in three dimensions around the center $R =0$) at $T=0$, and then reexpands to
future null infinity. The real amplitude
\begin{equation}\label{3.30}
  A(\omega) = Ce^{-a\omega}\ ,
\end{equation}
where $C$ and $a$ are constants, implies, by using Eq.\ (\ref{3.18}), solution
\begin{equation}\label{3.31}
  \phi = C\biggr\{\!\frac{[(a^2\!+\!R^2\!-\!T^2)^2 \!+ 4a^2T^2]^{1/2}\!+
  a^2\!+R^2\!-\!T^2}{(a^2+R^2-T^2)^2 + 4a^2T^2}\biggl\}^{\!\!\frac12}\,,
\end{equation}
regular everywhere. (Due to the factor $1/\sqrt{2}$ in Eq.\ (\ref{3.18})
$\phi$ here must be multiplied by $\sqrt{2}$ to get Eq.\ (3.15) in
\cite{ABSI}.) At spatial infinity, $R\rightarrow\infty$, $T$ fixed, we see
that
\begin{equation}\label{3.32}
  \phi = C\sqrt{2}\frac{1}{R} + O(1/R^2)\ ,
\end{equation}
in accordance with Eq.\ (\ref{3.29}). At the past null infinity
($R\rightarrow\infty$, $V=T+R$ fixed), we find
\begin{equation}\label{3.33}
  \phi = C\frac{\sqrt{2}}{2}\left[\frac{V +
  (V^2+a^2)^{1/2}}{V^2+a^2}\right]^{1/2} \frac{1}{\sqrt{R}} + O(1/R^{3/2})\ .
\end{equation}
(At future null infinity, $R\rightarrow\infty$, $U = T-R$ fixed, the same
expression, with $V$ replaced by $U$ follows.) A simple calculation, starting
from the formula (\ref{3.21}) for the profile $g(V)$ and using $A(\omega)$
from Eq.\ (\ref{3.30}), yields exactly the factor at $1/\sqrt{R}$ in Eq.\
(\ref{3.33}) (for integrals $\int_0^\infty dx\,e^{-ax}(1/\sqrt{x})\cos bx$ and
$\int_0^\infty dx\,e^{-ax}(1/\sqrt{x})\sin bx$ needed in the calculation, see
e.g., \cite{GR}, formulae 3.944, 13 and 14).

With $\phi$ given by Eq.\ (\ref{3.33}) one can find the explicit expression
for function $\gamma$ by solving Eqs.\ (\ref{3.7}) and (\ref{3.8}). It reads
\cite{WW}, \cite{ABSI} as follows:
\begin{multline}\label{3.34}
  \gamma = 4GC^2\left\{\frac{1}{a^2} - \frac{2R^2[(a^2+R^2-T^2)^2 -
  4a^2T^2]}{[(a^2+R^2-T^2)^2 + 4a^2T^2]^2} \right. \\ \left. +
  \frac{1}{a^2}\frac{R^2-a^2-T^2}{[(a^2+R^2-T^2)^2 + 4a^2T^2]^{1/2}}
  \right\}\ .
\end{multline}
The conicity at spatial infinity is thus given by
\begin{equation}\label{3.35}
  \gamma_\infty = 8G(C/a)^2\ .
\end{equation}
With this explicit solution one can verify directly relations
(\ref{3.15})--(\ref{3.17}) for the conicity as it is radiated ``in'' and
``out'', similarly to the Bondi mass in four dimensions.

\section{Hamiltonian formulation}
Let us essentially repeat the canonical treatment of Ashtekar and Pierri given
in \cite{AP} but with a slight modification in order to establish the
analogy with the model of Sec.\ 2. Let us start by considering the volume
part of the canonical action derived from the Einstein-Hilbert action by
assuming cylindrical symmetry in \cite{AP}:
\begin{equation}\label{AP action}
  S\! =\! \frac{1}{8G}\! \int\! dt \int^{\infty}_{0} \!dr \big( p_{\gamma} \,
  {\gamma}_{,t} + p_R \, R_{,t} + p_{\psi} \, {\psi}_{,t} - N \, C -  N^r \,
  C_r \big)\,,
\end{equation}
where
$$
  C = e^{-{\gamma}/2} \, \big( 2 \, R_{,rr} - {\gamma}_{,r} \, R_{,r}  -
  p_{\gamma} \, p_R + R^{-1} \, p_{\psi}^2/2 + R \, {\psi}_{,r}^2/2  \big)
$$
and
$$
  C_r = e^{-\gamma} \, \big( -2 \, p_{{\gamma},r} + p_{\gamma} \,
  {\gamma}_{,r} + p_{R} \, R_{,r} + p_{\psi} \, {\psi}_{,r} \big)
$$
are the constraint functions, $\gamma$, $R$, and $\psi$ are defined by Eq.\
(\ref{3.1}), $p_\gamma$, $p_R$ and $p_\psi$ are the conjugate momenta, while
$N$ and $N^r$ are Lagrange multipliers---the so-called lapse and shift
functions. One should add to this action the boundary energy term
\begin{equation}
- \frac{1}{4G} \, \int dt \, \Big( 1 - e^{-{\gamma}_{\infty}/2} \Big)\ ,
\label{AP energy}
\end{equation}
and specify the fall-off of $N$ according to $\lim_{r\rightarrow\infty}N = 1$
so that the action be differentiable \cite{AP}. However, this term is not
invariant under reparametrizations of the label time $t$ as there is no
temporal density present in the integrand in (\ref{AP energy}). Addition of
the bare term (\ref{AP energy}) to the action (\ref{AP action}) would imply a
privileged choice of asymptotic time. The total action would then not be in
the constraint-Hamiltonian form but rather in an already reduced form at
spatial infinity.

In order to recover the full constraint-Hamiltonian framework of our model in
Sec. 2, we need to justify the inclusion of a temporal density in (\ref{AP
  energy}). This can be done following the general approach by Beig and O'
Murchadha \cite{BO'M}. First, one considers fall-off conditions for the
configuration space data at $r \rightarrow \infty$. These have been specified
in \cite{AP}.  The configuration space fields approach infinity according to
\begin{equation}
\begin{gathered}
{\gamma}(t,r) \rightarrow {\gamma}_{\infty}(t) + O(1/r)\ ,\\
R(t,r) \rightarrow r \big( 1 + O(1/r) \big)\ ,\\
{\psi}(t,r) \rightarrow O(1/r) \ ,
\end{gathered}\label{falloff}
\end{equation}
where $r \, O(1/r)$, $r^2 \, O(1/r^2)$, etc. admit limits at $r \rightarrow
\infty$. (These limits generally depend on the time $t$.) One then requires
that the action of the Liouville form
\begin{equation}
\int^{\infty}_0 dr \, \Big( p_{\gamma} \, d{\gamma} + p_R \, dR + p_{\psi}
\, d{\psi} \Big)
\nonumber
\end{equation}
on vector fields of the form
\begin{equation}
\int^{\infty}_0 dr \, \Big( {\delta}{\gamma} \,
\frac{\partial}{{\partial}{\gamma}} + {\delta}R \,
\frac{\partial}{{\partial}R} + {\delta}{\psi} \,
\frac{\partial}{{\partial}{\psi}} \Big)
\nonumber
\end{equation}
should be finite. The resulting integral
\begin{equation}
\int^{\infty}_0 dr \, \Big( p_{\gamma} \,  {\delta}{\gamma}  + p_R \,
{\delta}R  + p_{\psi} \, {\delta}{\psi}  \Big)
\nonumber
\end{equation}
is finite if the momenta satisfy the following fall-off conditions at $r
\rightarrow \infty$:
\begin{equation}
\begin{gathered}
p_{\gamma}(t,r) \rightarrow O(1/r^2)\ ,\\
p_R(t,r) \rightarrow O(1/r^2)\ ,\\
p_{\psi}(t,r) \rightarrow O(1/r) \ .
\end{gathered}\label{falloff2}
\end{equation}

Next, concerning the behaviour of the lapse and shift, there are several
aspects that ought to be kept in mind. First, it is the finiteness and
differentiability of the Hamiltonian part
\begin{equation}\label{14a,1}
  {\mathcal H}_1[N,N^r] = \int_0^\infty dr\,(NC + N^rC_r)
\end{equation}
of action (\ref{AP action}) (cf.\ \cite{BO'M}). Second, ${\mathcal H}_1[N,N^r]$
has to generate a transformation within the phase space defined by the
boundary conditions (\ref{falloff}) and (\ref{falloff2}) \cite{BO'M}. Finally,
if we are going to have a full analogy to action (\ref{param action}) of Sec.\
2, we have to parametrize the model also at infinity, as it is done in
\cite{KSchw} for a spherically symmetric model.

The constraints functional (\ref{14a,1}) remains finite even if $N(r)$ and
$N^r(r)$ approach arbitrary temporal densities at $r \rightarrow \infty$;
namely,
\begin{equation}
\begin{gathered}
N(t,r) \rightarrow N_{\infty}(t) + O(1/r)\ ,\\
N^r(t,r) \rightarrow N^r_{\infty}(t) + O(1/r) \ ,
\end{gathered}\label{falloff3}
\end{equation}
where $N_{\infty}(t)$ has to be non-negative. The condition that the lapse and
shift should approach temporal densities at $r \rightarrow \infty$ is the
minimum requirement that is compatible with the invariance of the action
(\ref{AP action}) under reparametrizations of $t$.

Conditions (\ref{falloff})-(\ref{falloff3}) now imply that the action is not
differentiable. The problem comes from the variation of ${\gamma}(t,r)$
leading to the boundary term
\begin{eqnarray}
{\delta}S &\rightarrow& \frac{1}{8G} \, \int dt \int^{\infty}_0 dr \,
\Big( N \, e^{-{\gamma}/2} \, R_{,r} \, {\delta}{\gamma}  \Big)_{,r}
\nonumber
\\
&=& \frac{1}{8G} \, \int dt \, \Big( N_{\infty} \, e^{-{\gamma}_{\infty}/2}
\, {\delta}{\gamma}_{\infty}  \Big)
\label{bt}
\end{eqnarray}
at spatial infinity. In order to have a consistent canonical theory, one needs
to add to the action the boundary term
\begin{equation}
\frac{1}{4G}  \int dt \, \Big( N_{\infty} \, e^{-{\gamma}_{\infty}/2} \Big)\
,
\nonumber
\end{equation}
whose variation with respect to ${\gamma}(t,r)$ cancels the boundary term in
(\ref{bt}). The differentiable action is therefore
\begin{equation}
\begin{split}
S &= \frac{1}{8G} \, \int dt \int^{\infty}_0 dr \, \big( p_{\gamma} \,
{\gamma}_{,t} + p_R \, R_{,t} + p_{\psi} \, {\psi}_{,t} \\
&\mspace{220mu}- N \, C -  N^r \, C_r
\big)\\
&\quad - \frac{1}{4G} \, \int dt \, N_{\infty} \, \Big( 1 -
e^{-{\gamma}_{\infty}/2} \Big) \ .
\end{split}\label{AP ACTION}
\end{equation}
The boundary term in (\ref{AP ACTION}) has now been modified by the addition
of a constant in order that it coincides with the asymptotic energy derived
from first principles in \cite{AP}. There is no boundary term involving the
shift in spite of the fact that its asymptotic value may be non-zero. The
corresponding constraint functional generates an ``even supertranslation'' (in
the language of \cite{BO'M}) and is differentiable without any correction,
similarly to the case of four-dimensional general relativity, cf.\
\cite{BO'M}.

When varying the action (\ref{AP ACTION}) with respect to the lapse $N(t,r)$,
it is important to keep the ends of its variation fixed. Indeed, if
$N_{\infty}(t)$ is varied in (\ref{AP ACTION}) then one gets an unwanted field
equation implying that the asymptotic energy vanishes. It follows that the
action (\ref{AP ACTION}) is not yet in true constraint-Hamiltonian form.
Following Kucha\v{r} \cite{KSchw}, this can be improved by the
``parametrization at infinity'': $N_{\infty}(t)$ should be replaced by a
differentiated asymptotic time $dT_\infty/dt = T_{{\infty},t}(t)$. The
asymptotic time is determined by the asymptotic metric: it must hold $N_\infty
= 1$ if the parameter $t$ is chosen to be $T_\infty$. The time $T_{\infty}(t)$
can be varied in the ensuing action. Its variation leads to a redundant
equation amounting to the conservation of the asymptotic energy. One should
next introduce the momentum $P_{\infty}$ and add the associated constraint
(which is linear in $P_{\infty}$) to the action by a new Lagrange multiplier
$N_{\infty}(t)$.

In this way the action is brought into the true constraint-Hamiltonian form
\begin{multline}
S = \frac{1}{8G} \, \int dt \, \Big(  P_{\infty} \, T_{{\infty},t} \Big)
\\
 + \frac{1}{8G} \, \int dt \int^{\infty}_0 dr \, \Big( p_{\gamma} \,
{\gamma}_{,t} + p_R \, R_{,t} + p_{\psi} \, {\psi}_{,t}  \Big)
\\
 - \int dt \, N_{\infty} \, \Big( P_{\infty} + \frac{1}{4G} \, \big( 1-
 e^{-{\gamma}_{\infty}/2} \big) \Big)
\\
 - \frac{1}{8G} \, \int dt \int^{\infty}_0 dr \, \Big( N \, C +  N^r \, C_r
 \Big)  \, .
\label{AP param action}
\end{multline}
The multipliers $N$ and $N^r$ obey the asymptotic condition (\ref{falloff3}).
The action (\ref{AP param action}) is the analogue of the action (\ref{param
  action}) for the Newtonian particle. One can verify that the field equations
derived from the variations of (\ref{AP param action}) coincide with those of
Sec.\ 3, preserve the fall-off conditions (\ref{falloff})-(\ref{falloff3}) and
imply the conservation of the asymptotic energy.

Action (\ref{AP param action}) is our starting point for the canonical theory.
Although it corresponds to action (\ref{param action}) of Sec.\ 2, observe
that there is no a priori analogue of action (\ref{reduced action}) of Sec.\
2. We have to begin with the extended phase space ${\cal P}$ with coordinates
$\gamma(r)$, $p_{\gamma}(r)$, $R(r)$, $p_R(r)$, $\psi(r)$, $p_{\psi}(r)$,
$T_{\infty}$, $P_{\infty}$ and the symplectic form $\Omega_{\cal P}$,
\begin{multline*}
  \Omega_{\cal P} = \frac{1}{8G}\,dP_\infty\wedge dT_\infty +
  \frac{1}{8G} \int_0^\infty dr\, \bigl[dp_\gamma(r)\wedge d\gamma(r) \\
  +dp_R(r)\wedge dR(r) + dp_\psi(r)\wedge d\psi(r)\bigr]\ .
\end{multline*}
The constraint surface ${\mathcal C}$ is defined by ${\mathcal H}[N,N^r] = 0$
for all $N(r)$ and $N^r(r)$ satisfying the fall-off conditions, where
\begin{equation}
\begin{split}
{\mathcal H}[N,N^r] &= N_{\infty} \biggl( P_{\infty} + \frac{1}{4G} \, ( 1-
e^{-{\gamma}_{\infty}/2} ) \biggr) \\
&\qquad+ \frac{1}{8G} \, \int^{\infty}_0 dr \,
\bigl( N \, C + N^r \, C_r \bigr)\ .
\end{split}\label{HVF}
\end{equation}
The orbits are defined by the canonical action of the constraint functional
${\mathcal H}[N,N^r]$. The canonical transformations generated by (\ref{HVF})
are considered as a gauge transformation. Within the gauge group, there is no
distinction between the ``symmetry'' and the ``proper gauge'' as, e.g., in
\cite{AP}, \cite{BO'M} and \cite{honnef}. The functional (\ref{HVF}) generates
only reparametrizations both ``inside'' the spacetime {\em and} at infinity.
Symmetries are now generated by different functions. Indeed, the functional
(\ref{HVF}) has vanishing Poisson brackets with $P_\infty$ for any $N(r)$ and
$N^r(r)$ satisfying the conditions (\ref{falloff3}) because the variable
$\gamma_\infty$ is asymptotic value of the canonical coordinate $\gamma$ so
that $\{P_\infty,\gamma_\infty\} = 0$. Hence, it is the function $P_\infty$
that generates the symmetry. In general, we conjecture that one can introduce
privileged coordinates at infinity and that asymptotic symmetries are
generated by their conjugate momenta or suitable combinations of the momenta
and the coordinates (like, e.g., boosts).

The variable $T_\infty$ to which $P_\infty$ is conjugate is a kind of a
``privileged time'' but the surface $T_\infty =$ constant is neither a
transversal surface in the phase space, nor a Cauchy surface in each solution
spacetime. Indeed, the function $T_\infty -$ constant has vanishing Poisson
brackets with ${\mathcal H}[N,N^r]$ for all $N(r)$ and $N^r(r)$ whose
asymptotic values vanish; hence, the duly generalized regularity condition
(\ref{**}) is not satisfied. It follows that an infinite-dimensional
submanifold of each orbit lies in the surface $T_\infty =$ constant (Fig.\ 1).
This is connected to the fact that the condition $T_\infty =$ constant defines
only a particular section of infinity in each cylindrical wave spacetime but
not a Cauchy surface of the whole spacetime; there is a relation between
Cauchy and transversal surfaces, cf.\ \cite{HKij}.

\begin{figure}
\includegraphics[width=240pt]{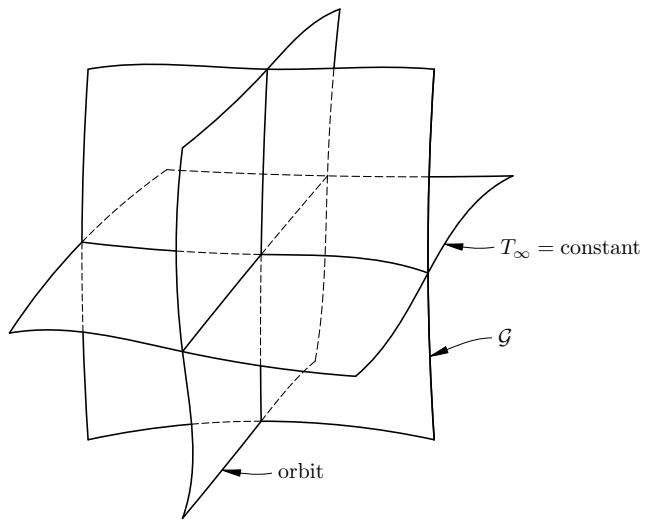}
\caption{Important surfaces in the constraint manifold $\mathcal C$.
  The intersection of $T_\infty = $ constant with any orbit is
  infinite-dimensional. The gauge condition surface $\mathcal G$ intersects
  each orbit in a (one-dimensional) dynamical trajectory of the reduced
  theory. The points common to $\mathcal G$ and each surface $T_\infty = $
  constant is a (infinite-dimensional) transversal surface
  ${\mathcal T}_{T_\infty}$.}
\end{figure}

The reduction by gauge condition, analogous to that described in Sec. 2.1,
starts by a choice of a one-dimensional family of transversal surfaces. Let us
denote the manifold formed by all chosen transversal surfaces in ${\mathcal
  C}$ by ${\mathcal G}$. In Sec.\ 2, a privileged choice of gauge has been
possible: ${\mathcal G}$ has been the family of surfaces $T = t$, $t \in
{\mathbb R}$, where $T$ is the privileged time. The nearest to this we can
come is to choose the transversal surfaces in ${\mathcal G}$ to be the
intersections of ${\mathcal G}$ and $T_\infty = $ constant (Fig.\ 1). There
are, of course, many choices of ${\mathcal G}$. One example of such a choice
is carried out in \cite{AP}.  Let us describe an analogous choice for our
action (\ref{AP param action}).

Following Ashtekar and Pierri, one may fix the part of the gauge associated
with the constraints $C(r)=0$ and $C_r(r)=0$ by imposing the gauge-fixing
conditions
\begin{equation}
R(r)=r\ ,\quad p_{\gamma}(r)=0 \, .
\label{AP fixing}
\end{equation}
These are the defining equations for ${\mathcal G}$. Viewed as constraints,
these conditions form together with the constraints $C(r)=0$ and $C_r(r)=0$ a
second-class system. The remaining constraint $P_{\infty} + \frac{1}{4G} \, (
1- e^{-{\gamma}_{\infty}/2} ) = 0$ can be taken care for by the gauge-fixing
condition
\begin{equation}
T_{\infty}= \text{constant}\ .
\label{AP fixing'}
\end{equation}
The surface ${\mathcal T}_{T_\infty}$ in ${\mathcal C}$ defined by (\ref{AP
  fixing})-(\ref{AP fixing'}) selects an initial datum from each gauge orbit
in $\cal C$. The gauge-condition surface ${\mathcal G}$ is swept by all
${\mathcal T}_{T_\infty}$.

In order to confirm that this reduction is admissible, let us add (\ref{AP
  fixing}) and (\ref{AP fixing'}) to the action (\ref{AP param action}) by
Lagrange multipliers $M$, $M^r$, and find out if the ensuing action determines
unique values for $N$, $N^r$. One obtains the action
\begin{equation}
\begin{split}
S &=\frac{1}{8G} \, \int dt \, \Bigl(  P_{\infty} \, T_{{\infty},t} \Bigr) \\
&\quad+\frac{1}{8G} \, \int dt \int^{\infty}_0 dr \, \Bigl( p_{\gamma} \,
{\gamma}_{,t} + p_R \, R_{,t} + p_{\psi} \, {\psi}_{,t}  \Bigr)\\
&\quad- \int dt \, N_{\infty} \, \Bigl( P_{\infty} + \frac{1}{4G} \,
\bigl( 1-e^{-{\gamma}_{\infty}/2} \bigr) \Bigr)\\
&\quad- \int dt \, M_{\infty} \,(T_{\infty}-t)\\
&\quad - \frac{1}{8G} \, \int dt \int^{\infty}_0 dr \, \Bigl( N \, C +  N^r \,C_r \Bigr)\\
&\quad- \int dt \int^{\infty}_0 dr \, \Bigl( M \, \bigl( R-r \bigr) + M^r \,p_{\gamma} \Bigr)  \ ,
\end{split}\label{second class action}
\end{equation}
where the set of conditions (\ref{AP fixing'}) is implemented by the
expression $\int dt \, M_{\infty} \, (T_{\infty}-t)$. Indeed, it is not
difficult to check that all redundant variables in (\ref{second class action})
can be expressed uniquely in terms of the canonical pair $(\psi,p_{\psi})$ by
solving the set of equations derived from the variation of these variables in
(\ref{second class action}). This confirms that the gauge-fixing conditions
(\ref{AP fixing}) are regular.  In particular, the unique reduced expressions
for the multipliers $N$, $N^r$ and $N_{\infty}$ are
\begin{align}
  &N(T_{\infty},R) =\notag\\
  &\quad={\rm exp}\bigg[\!-\frac{1}{4} \int^{\infty}_{R} dr
  \, \Big( {r}^{-1} p_{\psi}^2(T_{\infty},r) + r \,
  {\psi}_{,r}^2(T_{\infty},r)  \Big)    \bigg]\ ,\notag\\[6pt]
  &N^r(T_{\infty},R) = 0 \ ,\label{repsol1}
\end{align}
those for the canonical pairs $(\gamma,p_{\gamma})$, $(R,p_{R})$ and
$(T_{\infty},p_{T{\infty}})$ read
\begin{eqnarray*}
\gamma(T_{\infty},R) &=& \frac{1}{2} \, \int^{R}_{0} \, dr \, \Big( {r}^{-1}
\, p_{\psi}^2(T_{\infty},r) + r \, {\psi}_{,r}^2(T_{\infty},r)  \Big)\ ,
\\
p_{\gamma}(T_{\infty},R) &=& 0\ ,
\\
R(T_{\infty},r) &=& r\ ,
\\
p_{R}(T_{\infty},R) &=& - p_{\psi}(T_{\infty},R) \, {\psi}_{,R}(T_{\infty},R)\
,
\\
T_{\infty}(t) &=& t\ ,
\\
P_{\infty}(T_{\infty}) &=& - \frac{1}{4G} \, \bigg[ 1 - {\rm
  exp}\bigg(-\frac{1}{2}\gamma_\infty \bigg)\bigg] \ ,
\end{eqnarray*}
where
\begin{equation}\label{repsol2}
  \gamma_\infty = \frac{1}{2} \, \int^{\infty}_{0} \, dR \, \Big( {R}^{-1} \,
p_{\psi}^2(T_{\infty},R) + R \, {\psi}_{,R}^2(T_{\infty},R) \Big)\ ,
\end{equation}
and unique expressions also follow for the multipliers $M$, $M^r$,
$M_{\infty}$. The uniqueness of these expressions partially relies on the
conditions imposed on the canonical fields at $r=0$ (see, e.g., \cite{AP})
which force $\gamma(T_{\infty},0)$ to vanish for all $T_{\infty}$.

The reduced action for the remaining canonical pair
$\big(\psi(T_{\infty},R),p_{\psi}(T_{\infty},R)\big)$ on ${\mathcal G}
\subset {\cal C}$, parametrized by the values of the asymptotic time, is
therefore
\begin{multline}
  S = \frac{1}{8G} \, \int dT_{\infty} \int^{\infty}_0 dR \, \Big(
  p_{\psi}(T_{\infty},R) \, {\psi}_{,T_{\infty}}(T_{\infty},R)  \Big) \\
  -\frac{1}{4G} \, \int dT_{\infty} \, \Big( 1 -
  e^{-{\gamma}_{\infty}(T_{\infty})/2} \Big) \ ,
\label{AP reduced action}
\end{multline}
where ${\gamma}_{\infty}(T_{\infty})$ is expressed as a functional of
$\psi(T_{\infty},R)$ and $p_{\psi}(T_{\infty},R)\big)$ in (\ref{repsol2}). The
action (\ref{AP reduced action}) is analogous to the reduced action
(\ref{reduced action}). The phase space $\Gamma_1$ is described by
coordinates $\psi(R)$ and $p_\psi(R)$, while the symplectic form is
$$
  \Omega_1 = \int_0^\infty dR\,dp_\psi(R)\wedge d\psi(R)\ .
$$
The action (\ref{AP reduced action}) is precisely the reduced action of
Ashtekar and Pierri. In particular, the Ashtekar-Pierri time $t$ in Eq. (19)
of \cite{AP} corresponds to the time $T_{\infty}$ here.

Geometrically, $\psi(r)$, $p_\psi(r)$ and $T_\infty$ are coordinates on the
gauge-condition surface ${\mathcal G}$. The surfaces defined in ${\mathcal G}$
by the equation $T_\infty = $ constant are transversal surfaces in the {\em
  phase space} ${\mathcal P}$. However, they also determine a family of Cauchy
surfaces of constant Ashtekar--Pierri time in each solution {\em spacetime}
(cf.\ \cite{AP}). In this sense, the part (\ref{AP fixing}) of the gauge
condition determines a particular extension of the points at infinity defined
by $T_\infty = $ constant to whole Cauchy surfaces in the spacetimes. However,
different choices of ${\mathcal G}$ lead to different Cauchy surface
extensions of these points at infinity. Hence, two different choices of
${\mathcal G}$ entail two different choices of time so that the transformation
between the times has again the character of Eq.\ (\ref{8,1}) even if the part
(\ref{AP fixing'}) of gauge conditions remains always the same---only the
asymptotic values of these times have then to coincide. As noted at the end of
Sec.\ 2.1, it is the transformation (\ref{8,1}) between respective times which
causes difficulties in constructing a unique plausible quantum theory.

Considering the privileged symmetry generated by $P_\infty$, we can see that it
remains a symmetry of the reduced theory. It acts in $\mathcal G$ as follows,
\begin{equation}\label{17b,1}
  \Big(\psi(r), p_\psi(r), T_\infty\Big) \mapsto \Big(\psi(r), p_\psi(r),
  T_\infty + \tau\Big)\ ,
\end{equation}
while the original action of $P_\infty$ in ${\mathcal P}$ is
\begin{multline}\label{17b,2}
  \Big(\gamma(r), p_\gamma(r), R(r), p_R(r), \psi(r), p_\psi(r), T_\infty\Big)
  \\
  \mapsto
  \Big(\gamma(r), p_\gamma(r), R(r), p_R(r), \psi(r), p_\psi(r), T_\infty +
  \tau\Big)\ .
\end{multline}
The map (\ref{17b,2}) is {\em tangential} to $\mathcal G$ and the map
(\ref{17b,1}) is just the restriction of (\ref{17b,2}) to $\mathcal G$. This
follows from the fact that the constraints as well as relations (\ref{AP
  fixing}) that define $\mathcal G$ are independent of $T_\infty$.

The dynamics defined by action (\ref{AP reduced action}) determines a
foliation of $\mathcal G$ by one-dimension\-al dynamical trajectories
represented by two functions of two variables $\psi(R,T_\infty)$ and
$p_\psi(R,T_\infty)$. These are identical with the intersections of $\mathcal
G$ with the orbits. In this way, we obtain a bijection between integrals of
motion of the reduced theory and Dirac observables. On one hand, any Dirac
observable is constant along each orbit.  Hence, it must also be constant
along each dynamical trajectory of action (\ref{AP reduced action}). On the
other, any function on $\mathcal G$ that is constant along each dynamical
trajectory defines a unique extension to $\mathcal C$ that is constant along
each orbit.

This relation between the Dirac observables of the extended system and the
integrals of motion of the reduced theory, together with the compatibility of
the symmetry groups generated by $P_\infty$ in the extended and reduced
theories, justify the approach of Secs.\ 5 and 6, where we shall construct the
gauge-invariant dynamics starting from the gauge-dependent action (\ref{AP
  reduced action}).

\section{Physical phase space $\Gamma_2$}
In this section, we choose a complete set of Dirac observables, find their
Poisson algebra and calculate their Poisson brackets with the symmetry
generator $P_\infty$. This task is simplified if we start from
Ashtekar--Pierri reduced action (\ref{AP reduced action}) instead of the
original parametrized action (\ref{AP param action}). According to what has
been shown in the previous sections, the result is independent of the gauge
chosen to reduce the action (\ref{AP param action}).

The reduced action (\ref{AP reduced action}) can be rewritten in terms of the
rescaled field $\phi=\psi/{\sqrt{8G}}$ introduced in Sec.\ 3 as follows:
$$
  S = \int dt dR\, (\pi_\phi\dot{\phi} - H)\ ,
$$
where
\begin{equation}\label{1-4}
  \gamma_\infty = 4G\int_0^\infty dR\,\left(\frac{1}{R}\pi_\phi^2 +
  R\phi^{\prime 2}\right)
\end{equation}
enters the Hamiltonian
\begin{equation}\label{1-5}
  H = \frac{1}{4G}\left(1-e^{-\frac{1}{2}\gamma_\infty}\right)\ .
\end{equation}
For simplicity, the notation for our time $T_{\infty}$ has been changed to the
Ashtekar-Pierri notation $t$. The Hamiltonian depends on $t$ only through
$\pi_\phi$ and $\phi$ so that $H$ and $\gamma_\infty$ are constants of motion,
\begin{equation}\label{1-3}
 \dot{\gamma}_\infty = 0\ .
\end{equation}

The canonical equations that follow from the action are
\begin{eqnarray}\label{1-1}
  \dot{\pi}_\phi &=& e^{-\frac{1}{2}\gamma_\infty}  (R\phi')' \ , \\
\label{1-2}
  \dot{\phi} &=& e^{-\frac{1}{2}\gamma_\infty} \frac{1}{R}\pi_\phi \ .
\end{eqnarray}
Eqs.\ (\ref{1-2}) and (\ref{1-3}) imply
$$
  \ddot{\phi} = e^{-\frac{1}{2}\gamma_\infty}\frac{1}{R}\dot{\pi}_\phi
$$
so that
\begin{equation}\label{5,*}
  e^{\gamma_\infty}\ddot{\phi} = \frac{\partial^2\phi}{\partial R^2} +
  \frac{1}{R}\frac{\partial\phi}{\partial R}\ .
\end{equation}
If we use the relation between the Einstein-Rosen time $T$ and the
Ashtekar-Pierri time $t$ (see \cite{AP}),
\begin{equation}\label{2-1}
  T(t) = e^{-\frac{1}{2}\gamma_\infty}t\ ,
\end{equation}
then Eq.\ (\ref{5,*}) becomes the wave equation (\ref{3.6}).
The general solution to Eq.\ (\ref{3.6}) is given by Eq.\ (\ref{3.18}), which
can be written in terms of time $t$ as
\begin{equation}\label{2-2}
  \phi(t,R) = \frac{1}{\sqrt{2}}\int_0^\infty
  d\omega\,\left[A(\omega)J_0(\omega R)e^{-i\omega T(t)} +\cxconj\right]\ ,
\end{equation}
and Eq.\ (\ref{1-2}) yields
\begin{equation}\label{2-3}
  \pi_{\phi}(t,R) = \frac{R}{\sqrt{2}}\int_0^\infty
  d\omega\,\left[-i\omega A(\omega)J_0(\omega R)e^{-i\omega T(t)} +
  \cxconj\right]\ .
\end{equation}
Eqs.\ (\ref{2-2}) and (\ref{2-3}) describe the general solution to the
canonical equations (\ref{1-1}) and (\ref{1-2}) in terms of the set of
constants $A(\omega)$. Hence, the parameters $A(\omega)$ can serve as
coordinates on the physical phase space $\Gamma_2$.

The physical phase space is a symplectic manifold. Its full structure can be
obtained if we find a transversal surface. As has been explained in Sec.\ 2,
any transversal surface, together with the symplectic form that results from
pulling back the symplectic form from the extended phase space to the
transversal surface, form the structure that is isomorphic to the physical
phase space.  In our case, the initial data $\phi_0$ and $\pi_{\phi 0}$ of the
canonical coordinates $\phi$ and $\pi_{\phi}$ at the Cauchy surface $t=0$
determine a unique solution (\ref{2-2}) and (\ref{2-3}) so that they can also
be considered as coordinates on the physical phase space $\Gamma_1$. Moreover,
the surface ${\mathcal T}_{0}$ defined by the Ashtekar and Pierri gauge
(\ref{AP fixing}) together with the condition $T_\infty=0$ {\em is} a
transversal surface. Hence, the symplectic form $\Omega$ on the physical phase
space with respect to the coordinates $\phi_0$ and $\pi_{\phi 0}$ is
\begin{equation}\label{3-1}
  \Omega_2 = \int_0^\infty dR\, d\pi_{\phi 0}(R)\wedge d\phi_0(R)
\end{equation}
because this is the pull back of $\Omega_{\mathcal P}$ to ${\mathcal T}_{0}$
by the injection map of ${\mathcal T}_{0}$ into $\mathcal P$; the manifold
${\mathcal T}_{0}$ with this symplectic form is isomorphic to the physical
phase space $\Gamma_2$.

The relations between the parameters $A(\omega)$ and $\phi_0$, $\pi_{\phi 0}$
can be obtained from Eqs.\ (\ref{2-2}) and (\ref{2-3}):
\begin{equation}\label{3-2}
  \phi_0(R) = \frac{1}{\sqrt{2}}\int_0^\infty
  d\omega\,J_0(\omega R)[A(\omega) + A^*(\omega)]\ ,
\end{equation}
and
\begin{equation}\label{3-3}
  \pi_{\phi 0}(R) = -\frac{iR}{\sqrt{2}}\int_0^\infty
  d\omega\,\omega J_0(\omega R)[A(\omega)- A^*(\omega)]\ ,
\end{equation}
while the inverse transformation is analogous to Eq.\ (\ref{3.28}):
\begin{equation}\label{3-4}
  A(\omega) =  \frac{1}{\sqrt{2}}\int_0^\infty dR\,J_0(\omega R)[\omega
  R\phi_0(R) - i\pi_{\phi 0}(R)]\ .
\end{equation}

A further set of parameters to determine points of the physical phase space
are the ${\mathcal I}^-$ null data $g(V)$ or ${\mathcal I}^+$ null data
$f(U)$. The transformations between $A(\omega)$ and $g(V)$ is given by Eqs.\
(\ref{3.21}) and (\ref{3.25}), those between $A(\omega)$ and $f(V)$ by Eqs.\
(\ref{3.20}) and (\ref{3.27}).

The quantity $\gamma_\infty$ is a function on the physical phase space given,
in terms of the four different coordinate systems, by Eqs.\
(\ref{1-4}), (\ref{3.16}) and (\ref{3.17}). Eq.\ (\ref{1-4}), into which Eqs.\
(\ref{3-2}) and (\ref{3-3}) are substituted, yields after some simple
transformations the expression for $\gamma_\infty$ in terms of $A(\omega)$:
\begin{equation}\label{4-1}
  \gamma_\infty = 8G\int_0^\infty d\omega\,\omega
  A^*(\omega)A(\omega) \ .
\end{equation}

We can also express the symplectic form (\ref{3-1}) in terms of
$A(\omega)$. If  Eqs.\ (\ref{3-2}) and (\ref{3-3}) are
substituted into Eq.\ (\ref{3-1}), we obtain
\begin{equation*}
\begin{split}
  \Omega_2
  = -\frac{i}{2}\int_0^\infty &d\omega\int_0^\infty d\omega'\int_0^\infty
  dR\,\omega'RJ_0(\omega R)J_0(\omega' R) \\
  \times\bigl[\!&-dA(\omega)\wedge dA(\omega')
  + dA(\omega)\wedge dA^*(\omega') \\
  &- dA^*(\omega)\wedge dA(\omega') +
  dA^*(\omega)\wedge dA^*(\omega')\bigr]\ .
\end{split}
\end{equation*}
The formulae (\ref{HT1}) and (\ref{HT2}) imply, however, that
\begin{equation}\label{5-1}
  \int_0^\infty dR\,RJ_0(\omega R)J_0(\omega' R) =
  \frac{1}{\sqrt{\omega\omega'}}\delta(\omega-\omega') \ .
\end{equation}
Hence, using the antisymmetry of the wedge product, we obtain finally
\begin{equation}\label{5-2}
  \Omega_2 = i\int_0^\infty
  d\omega\,dA^*(\omega)\wedge dA(\omega) \ .
\end{equation}

Let us also express the symplectic form of the physical phase space in terms of
the asymptotic null data $g(V)$ and $f(U)$. Eqs.\ (\ref{5-2}) and (\ref{3.25})
give
\begin{multline*}
  i\int_0^\infty d\omega\,dA^*(\omega)\wedge dA(\omega) =\\ =
  -\frac{i}{\pi}\int_{-\infty}^\infty \!dV\int_{-\infty}^\infty \!d\bar{V}\,
  dg(V)\wedge dg(\bar{V})\int_0^\infty \!d\omega\,\omega e^{i\omega(V-\bar{V})}
  \ .
\end{multline*}
Since the wedge product is antisymmetric in $V$ and $V'$, only the
antisymmetric part of the integral over $\omega$ contributes to the
result. However,
\begin{multline*}
  \frac{i}{2\pi}\int_0^\infty d\omega\,\omega\left[e^{i\omega(V-\bar{V})} -
  e^{-i\omega(V-\bar{V})}\right] \\ =
  \frac{1}{2\pi}\int_{-\infty}^\infty d\omega\,i\omega e^{i\omega(V-\bar{V})} =
  \frac{1}{2\pi}\frac{d}{dV}\int_{-\infty}^\infty d\omega\,
  e^{i\omega(V-\bar{V})} \\ =  \frac{d}{dV}\delta(V-\bar{V})\ .
\end{multline*}
Hence,
\begin{equation}\label{6-1}
  \Omega_2 = \int_{-\infty}^\infty dV\, dg'(V)\wedge dg(V) \ .
\end{equation}
By analogous calculation, Eqs.\ (\ref{5-2}) and (\ref{3.27}) yield
\begin{equation}\label{6-2}
  \Omega_2 = \int_{-\infty}^\infty
  dU\, df'(U)\wedge df(U) \ .
\end{equation}

Finally, let us calculate the transformation between $f(U)$ and $g(V)$ if
$f(U)$ is defined by the solution determined by $g(V)$. Such a transformation
is, therefore, entailed by Eqs.\ (\ref{3.20}) and (\ref{3.25}):
$$
  f(U) = \frac{1}{2\pi}\int_{-\infty}^\infty \!\!dV\, g(V)\!\int_0^\infty
  \!\!d\omega\Bigl[ie^{i\omega(U-V)} \!-\! ie^{-i\omega(U-V)}\Bigr]\,.
$$
The distribution $D(U-V)$ defined by the integral over $\omega$ can be
approximated by a convergent series of distributions $D_\epsilon(U-V)$ (see
\cite{GS}),
$$
  \lim_{\epsilon\rightarrow 0}D_\epsilon(U-V) = D(U-V)\ ,
$$
where $\epsilon> 0$ and
\begin{equation*}
\begin{split}
  D_\epsilon(U-V) &= \int_0^\infty
  d\omega\,\left[ie^{i\omega(U-V)-\omega\epsilon} -
  ie^{-i\omega(U-V)-\omega\epsilon}\right]\\
  &= -2\frac{U-V}{(U-V)^2 +\epsilon^2} \ .
\end{split}
\end{equation*}
However,
\begin{equation*}
\begin{split}
 &\lim_{\epsilon\rightarrow 0}\int_{-\infty}^\infty
 dV\,\left[-2\frac{U-V}{(U-V)^2 +\epsilon^2}\right] g(V) =\\
 &\mspace{200mu}= -2P\int_{-\infty}^\infty dV\,\frac{g(V)}{U-V}\ ,
\end{split}
\end{equation*}
where $P$ denotes the principal value. Hence,
\begin{equation}\label{8-1}
  f(U) = -\frac{1}{\pi}P\int_{-\infty}^\infty dV\,\frac{g(V)}{U-V} \ .
\end{equation}

\section{Representation of symmetries in the physical phase space}
There are two interesting symmetries to be represented in the physical phase
space. The first is the infinitesimal time translation, and the second is the
map
$$
  \sigma : {\mathcal I}^- \mapsto {\mathcal I}^+\ ,
$$
defined by $U = V$ in terms of coordinates $U$ at ${\mathcal I}^+$ and $V$ at
${\mathcal I}^-$ (an analogous symmetry transformation has been studied in
\cite{HI}).

The push-forward action of the infinitesimal translation $t \mapsto t +\delta
t$ on the solution fields $\phi(t,R)$ and $\pi_{\phi}(t,R)$ is given by
\begin{eqnarray}\label{8-2}
  \phi(t,R) &\mapsto & \phi(t-\delta t,R)\ , \\
\label{8-3}
  \pi_{\phi}(t,R)) &\mapsto & \pi_{\phi}(t-\delta t,R)\ .
\end{eqnarray}
Substituting $t-\delta t$ for $t$ into Eqs.\ (\ref{2-2}) and (\ref{2-3}) and
comparing the results with these equations leads to $A(\omega)\mapsto
A(\omega)+ \delta A(\omega)$, where
\begin{equation}\label{9-1}
  \delta A(\omega) = i\omega e^{-\frac{1}{2}\gamma_\infty}A(\omega)\delta t \ .
\end{equation}
The same result can be obtained, if we put $\phi_0(R) + \delta\phi_0(R)$ and
$\pi_{\phi 0}(R) + \delta\pi_{\phi 0}(R)$ into Eq.\ (\ref{3-4}) and
calculate the corresponding $\delta A(\omega)$. We must utilize the fact that
$$
  \delta\phi_0(R) = -\dot{\phi}_0(R)\delta t\ ,\quad \delta\pi_{\phi 0}(R) =
  -\dot{\pi}_{\phi 0}(R)\delta t\ ,
$$
express the time derivatives with the help of the equations of motion
(\ref{1-1}) and (\ref{1-2}), transfer the $r$-derivatives from $\phi_0(R)$ to
$J_0(\omega R)$, and use the Bessel equation that is satisfied by $J_0(\omega
R)$,
$$
  -\frac{1}{R}(RJ'_0(\omega R))' = \omega^2J_0(\omega R)\ .
$$

It follows that the action of the infinitesimal time translation is
canonically generated  by the function $-H$ defined by Eq.\ (\ref{1-5}), with
$\gamma_\infty$ given by Eq.\ (\ref{1-4}), where $\phi$ and $\pi_\phi$ are
replaced by $\phi_0$ and $\pi_{\phi 0}$. (Indeed, the momentum conjugate to
$t$ is $P_\infty = -H$.) We thus have
\begin{equation*}
\begin{aligned}
  \delta\phi_0(R) &= \{\phi_0(R),-H\} \ ,\\
  \delta\pi_{\phi 0}(R) &= \{\pi_{\phi 0}(R),-H\} \ ,\\
  \delta A(\omega) &= \{A(\omega),-H\} \ ,
\end{aligned}
\end{equation*}
and obtain analogously
\begin{eqnarray}
\label{10-2}
  \delta f(U) &=& \{f(U),-H\} \ ,\\
\label{10-3}
  \delta g(V) &=& \{g(V),-H\} \ .
\end{eqnarray}

Let us express $\delta f(U)$ and $\delta g(V)$ explicitly from the action of
translations. Since the whole solution is shifted along the background
manifold defined by the coordinates $t$ and $R$ by $t \mapsto t + \delta t$,
$R \mapsto R$, we have, regarding Eq.\ (\ref{2-1}),
$$
  U = T-R \mapsto e^{-\frac{1}{2}\gamma_\infty}(t + \delta t) - R = U +
  e^{-\frac{1}{2}\gamma_\infty}\delta t\ ,
$$
and similarly for $V$:
$$
  V \mapsto V + e^{-\frac{1}{2}\gamma_\infty}\delta t\ .
$$
Hence,
\begin{eqnarray}\label{10-4}
   \delta f(U) &=& -f'(U)e^{-\frac{1}{2}\gamma_\infty}\delta t\ , \\
\label{10-5}
   \delta g(V) &=& -g'(V)e^{-\frac{1}{2}\gamma_\infty}\delta t\ .
\end{eqnarray}
The same relations result from the Poisson brackets (\ref{10-2}) and
(\ref{10-3}), if Eqs.\ (\ref{3.20}), (\ref{3.21}), (\ref{1-5}), (\ref{4-1})
and (\ref{5-2}) are used; notice that Eq.\ (\ref{5-2}) implies
\begin{equation}\label{11-1}
  \{A(\omega),A(\omega')\} = 0 \ ,\quad \{A^*(\omega),A^*(\omega')\} = 0 \ ,
\end{equation}
and
\begin{equation}\label{11-2}
  \{A(\omega),A^*(\omega')\} = -i\delta(\omega - \omega') \ .
\end{equation}

As has been explained in Sec.\ 2 (see also \cite{honnef}), the dynamics of the
Dirac observables is defined by the comparison of the equations of motion with
the action of symmetry. Since the equations of motion for the Dirac
observables are trivial (the observables remain constant), the symmetry action
alone gives the total dynamical change.

The second symmetry $\sigma : {\mathcal I}^- \mapsto {\mathcal I}^+$ is a
purely asymptotic one, similarly to $T_\infty \mapsto T_\infty + \tau$. Its
action on solutions can be found in an analogous way via the Cauchy data for
solutions at $\mathcal I$'s. We consider Cauchy null datum $g_1(V)$ at
${\mathcal I}^-$ as defining solution $\phi_1(t,R)$.  Then we push forward the
field $g_1(V)$ at ${\mathcal I}^-$ to ${\mathcal I}^+$ by $\sigma_*$, which
results in Cauchy datum $f_2(U)$ at ${\mathcal I}^+$. The Cauchy datum
$f_2(U)$ determines another solution $\phi_2(t,R)$, and we define it as the
image of $\phi_1(t,R)$ by $\sigma$. The corresponding map in $\Gamma_2$ can be
calculated by using coordinates $f(U)$ in $\Gamma_2$. Solution $\phi_2(t,R)$
has coordinate $f_2(U)$, and let $\phi_1(t,R)$ has coordinate $f_1(U)$. Then
the point $f_2(U)$ in $\Gamma_2$ is the image of the point $f_1(U)$ of
$\Gamma_2$ by map $\sigma$. The dynamics defined by $\sigma$ is the inverse
map because it compares the evolution by the wave equation (which is trivial
because Dirac observables remain constant) with the map by $\sigma$ (cf.\
\cite{HI}).

The push forward map $\sigma_*$ of fields at ${\mathcal I}^-$ to those at
${\mathcal I}^+$ acts as follows:
$$
  \sigma_*g(V) = f(U)\ ,
$$
where $f(U) = g(U)$. It follows immediately that the dynamical evolution
defined by the ``zero motion'' $\sigma$ is represented by transformation
(\ref{8-1}).

We can also introduce Fourier amplitudes $a(\omega)$ and $b(\omega)$ of the
asymptotic data by
$$
  a(\omega) = A(\omega)e^{i\pi/4}\ ,\quad b(\omega) = A(\omega)e^{-i\pi/4}\ ,
$$
so that Eqs.\ (\ref{3.20}) and (\ref{3.21}) become
\begin{eqnarray*}
  f(U) &=& \frac{1}{2\sqrt{\pi}}\int_0^\infty
  \frac{d\omega}{\sqrt{\omega}}\,\left[b(\omega)e^{-i\omega U} +
  \cxconj\right] \ , \\
  g(V) &=& \frac{1}{2\sqrt{\pi}}\int_0^\infty
  \frac{d\omega}{\sqrt{\omega}}\,\left[a(\omega)e^{-i\omega V} +
  \cxconj\right] \ .
\end{eqnarray*}
The push forward of the amplitudes is clearly given by
$$
  \sigma_*a(\omega) = b(\omega)\ ,
$$
where $b(\omega) = a(\omega)$. Canonical representation of  $\sigma$ is,
therefore:
\begin{equation}\label{12-1}
  b(\omega) = a(\omega)e^{-i\pi/2} = -ia(\omega) \ .
\end{equation}
This map (or (\ref{8-1})) becomes the $S$-matrix of the one-particle sector in
the quantum theory of the model.

\section{Quantum theory}
It is easy to construct the Hilbert space, the operators representing the
Dirac observables, the Hamiltonian, and to define the scattering matrix in the
standard way of quantization of linear field theories (see, e.g., \cite{DW1}
or \cite{wald}). A sketch thereof will be described in this section.

Let us start from the Poisson brackets (\ref{11-1}) and (\ref{11-2}) for the
observables $A(\omega)$. Roughly, in the canonical quantization, Poisson
brackets are replaced by commutators multiplied by $i$ (the units are chosen
so that the Planck constant is 1). Then, we have
\begin{equation}\label{1:1}
  [\hat{A}(\omega'),\hat{A}(\omega)] = 0\ ,\quad
  [\hat{A}(\omega'),\hat{A}^\dagger(\omega)] = \delta(\omega' - \omega)\ .
\end{equation}
These are commutators of the {\em annihilation} and {\em creation operators}
of a quantum field theory for a continuous spectrum. They form our starting
point.

For many constructions it is favorable to use a smeared version of the
operators. We choose any complete orthonormal basis of (complex) functions
$X_n(\omega)$, where $\omega \in (0,\infty)$. This means that any complex
function$f$ can be decomposed,
$$
  f(\omega) = \sum_n f_n X_n(\omega)\ ,
$$
where $f_n$ are complex coefficients, and that
\begin{equation}\label{2:1}
  \int_0^\infty d\omega\,X^*_n(\omega)X_m(\omega) = \delta_{nm} \ .
\end{equation}
Defining
\begin{equation}\label{2:2}
  \hat{A}_n = \int_0^\infty d\omega\,X^*_n(\omega)\hat{A}(\omega) \ ,
\end{equation}
we obtain
\begin{equation}\label{2:3}
  \hat{A}(\omega) = \sum_n X_n(\omega)\hat{A}_n\ ,
\end{equation}
and
\begin{equation}\label{2:4}
  [\hat{A}_n,\hat{A}_m] = 0\ ,\quad
  [\hat{A}_n,\hat{A}^\dagger_m] = \delta_{nm} \ .
\end{equation}
Then we can define the vacuum state $|0\rangle$ by
\begin{equation}\label{2:5}
  \hat{A}_n|0\rangle = 0 \quad \forall n\ ,\quad \langle0|0\rangle = 1 \ ,
\end{equation}
which also implies that
\begin{equation}\label{3:1}
  \hat{A}(\omega)|0\rangle = 0\quad \forall \omega \ .
\end{equation}
The elements of a complete basis in the Hilbert space are obtained by
application of any number of creation operators $\hat{A}^\dagger_m$ to
$|0\rangle$; if the total number of the creation operators is $N$, then the
state is an $N$-graviton state. The scalar product is defined by scalar
products of the basis elements, which, in turn, are determined by the
commutation rules (\ref{2:4}) and the conditions (\ref{2:5}). For example,
\begin{equation*}
\begin{split}
  \left(\hat{A}^\dagger_m|0\rangle,\hat{A}^\dagger_n|0\rangle\right) &=
  \langle0|\hat{A}_m\hat{A}^\dagger_n|0\rangle \\
  &= \langle 0|\hat{A}_m^\dagger\hat{A}_n + \delta_{nm}|0\rangle =
  \delta_{nm}\ .
\end{split}
\end{equation*}
The Hilbert space defined in this way is often called the {\em Fock space} and
we denote it by $\mathcal F$.

Those Dirac observables defined in Sec.\ 5 that are linear in the variables
$A(\omega)$ and $A^\dagger(\omega)$ can be associated with operators on
$\mathcal F$ that are linear combinations of the operators $\hat{A}(\omega)$
and $\hat{A}^\dagger(\omega)$ with the same coefficients. This definition
preserves the relation between Poisson brackets and commutators. For example,
we define
\begin{equation*}
\begin{split}
\hat{f}(U) =
\frac{1}{2\sqrt{\pi}}\int_0^\infty
\frac{d\omega}{\sqrt{\omega}}\,
&\Bigl[\hat{A}(\omega)e^{-i(\pi/4) -i\omega U} \\
&\qquad+\hat{A}^\dagger(\omega)e^{i(\pi/4) +i\omega U}\Bigr] \ .
\end{split}
\end{equation*}
The matrix elements of $\hat{f}(U)$ with respect to the Fock basis are easily
calculated by using the decomposition (\ref{2:3}). In such a way, we have a
Hilbert space and the operators that correspond to the basic quantities.

In order to construct the Hamiltonian, we start from Eqs.\ (\ref{1-5}) and
(\ref{4-1}). We define the quadratic operator $\hat{\gamma}_\infty$ by the
normal factor ordering:
$$
  \hat{\gamma}_\infty = 8G\int_0^\infty
  d\omega\,\omega\hat{A}^\dagger(\omega)\hat{A}(\omega) =
  \sum_{nm}\omega_{nm}\hat{A}_n^\dagger\hat{A}_m\ ,
$$
where
$$
  \omega_{nm} = 8G\int_0^\infty
  d\omega\,\omega X^*_n(\omega) X_m(\omega)\ .
$$
Then $\hat{\gamma}_\infty|0\rangle = 0$. The operator $\hat{\gamma}_\infty$ is
self-adjoint on $\mathcal F$; it has a continuous spectrum. Its (generalized)
eigenvectors form a $\delta$-normalized basis of $\mathcal F$, elements of
which are obtained from the vacuum by application of any number of the
creation operators $\hat{A}^\dagger(\omega)$ (and a normalization factor). For
example,
$$
  \hat{\gamma}_\infty\left(\hat{A}^\dagger(\omega)|0\rangle\right) =
  8G\omega\left(\hat{A}^\dagger(\omega)|0\rangle\right)\ .
$$
Then, any function of $\hat{\gamma}_\infty$ can be defined by the spectral
theorem (see, e.g., \cite{D}): it has the same eigenvectors, and its
eigenvalues are the values that the function has on the corresponding
eigenvalues of $\hat{\gamma}_\infty$. In this way, the Hamilton operator
$$
  \hat{H} = \frac{1}{4G}\left[1 -
  \exp\left(-\frac{1}{2}\hat{\gamma}_\infty\right)\right]
$$
is well-defined. For example,
$$
  \hat{H}|0\rangle = \frac{1}{4G}\left[1 -
  \exp\left(-\frac{1}{2}\times 0\right)\right] = 0\ ,
$$
because $\hat{\gamma}_\infty$ has the
eigenvalue zero on $|0\rangle$.

Finally, we can define the scattering matrix $\hat{S}$. In order to do that,
we have to determine what are the in- and out-states. It seems natural to take
the states that result
\pagebreak[2]
from applying any number of the operators
$\hat{a}^\dagger(\omega)$ to $|0\rangle$ corresponding to the observables
$a(\omega)$ of Sec.\ 6 as the in-states. Similarly, the out-states can be
defined by $b(\omega)$. From Eq.\ (\ref{12-1}), we have a simple Bogolyubov
transformation between $\hat{a}(\omega)$ and $\hat{b}(\omega)$:
$$
  \hat{a}(\omega) = i\hat{b}(\omega)\ .
$$
The construction of the scattering matrix that implements a given Bogolyubov
transformation is described in \cite{wald} or \cite{DW1}. We shall skip it
because it lies outside the scope of this paper.

Another interesting question is, what is the relation between the Hamiltonian
$\hat{H}$ and the scattering operator $\hat{S}$. There are methods of
calculating $\hat{S}$ from $\hat{H}$: one has to take some limits within the
Euclidean regime (see, e.g., \cite{DW2}). However, an application, or even an
applicability, of these methods to our case also lies outside the scope of
this work.

\begin{acknowledgments}
J.B. acknowledges
the hospitality of the Institute of Theoretical Physics, University of Berne,
and a partial support from the grant GA\v{C}R 202/02/0735 of the Czech
Republic. This work was also supported by the Swiss National Science
Foundation and the Tomalla Foundation Zurich. We thank Miroslav Bro\v{z} and
Pavel Krtou\v{s} for the help with the manuscript.
\end{acknowledgments}

\end{document}